\documentclass[article,lineno]{biometrika}

\usepackage[utf8]{inputenc}
\usepackage{natbib}
\usepackage{amsfonts} 
\usepackage{bm}
\usepackage{xcolor}
 \usepackage{mathtools}
 \usepackage{ mathdots }
 \usepackage{amsmath}
\usepackage{dsfont}
\usepackage{relsize}
\usepackage{enumitem}
\usepackage{comment}
\usepackage[font={small,it}]{caption}
\usepackage{multirow}
\usepackage{float}
\usepackage{listings}
\usepackage{subcaption}
\usepackage{appendix}
\usepackage{graphics}
\usepackage{chngcntr}
\usepackage{hyperref}

\nolinenumbers

\usepackage{newtxtext}
\usepackage[subscriptcorrection]{newtxmath}

\graphicspath{{./art/}}

\usepackage[plain,noend]{algorithm2e}

\makeatletter
\renewcommand{\algocf@captiontext}[2]{#1\algocf@typo. \AlCapFnt{}#2} 
\def\@algocf@capt@plain{top}
\renewcommand{\algocf@makecaption}[2]{%
  \addtolength{\hsize}{\algomargin}%
  \sbox\@tempboxa{\algocf@captiontext{#1}{#2}}%
  \ifdim\wd\@tempboxa >\hsize
    \hskip .5\algomargin%
    \parbox[t]{\hsize}{\algocf@captiontext{#1}{#2}}
  \else%
    \global\@minipagefalse%
    \hbox to\hsize{\box\@tempboxa}
  \fi%
  \addtolength{\hsize}{-\algomargin}%
}
\makeatother

\begin{document}

\jname{}
\jyear{}
\jvol{}
\jnum{}
\cyear{}


\markboth{F. Asgari et~al.}{Functional structural equation modeling}

\title{Functional structural equation modeling with latent variables}

\author{Fatemeh Asgari$^{*1,2,3}$,Valeria Vitelli$^{\dagger2}$, and Uta Sailer$^{\dagger\dagger1}$}
\affil{$^1$Department of Behavioural Medicine, Institute of Basic Medical Sciences, Faculty of Medicine, University of Oslo, Oslo, Norway\\$^2$Oslo Centre for Biostatistics and Epidemiology, Department of Biostatistics, Institute of Basic Medical Sciences, University of Oslo, Oslo, Norway\\$^3$Department of Endocrinology, Obesity and Nutrition, Vestfold Hospital Trust, Tønsberg, Norway
\email{fatemeh.asgari@medisin.uio.no}$^*$
\email{valeria.vitelli@medisin.uio.no}$^\dagger$
\email{uta.sailer@medisin.uio.no}$^{\dagger\dagger}$
}


\maketitle

\begin{abstract}
Handling latent variables in Structural Equation Models (SEMs) in a case where both the latent variables and their corresponding indicators in the measurement error part of the model are random curves presents significant challenges, especially with sparse data. In this paper, we develop a novel family of Functional Structural Equation Models (FSEMs) that incorporate latent variables modeled as Gaussian Processes (GPs). The introduced FSEMs are built upon functional regression models having response variables modeled as underlying GPs. The model flexibly adapts to cases when the random curves' realizations are observed only over a sparse subset of the domain, and the inferential framework is based on a restricted maximum likelihood approach. The advantage of this framework lies in its ability and flexibility in handling various data scenarios, including regularly and irregularly spaced points and thus missing data. To extract smooth estimates for the functional parameters, we employ a penalized likelihood approach that selects the smoothing parameters using a cross-validation method.   We evaluate the performance of the proposed model using simulation studies and a real data example, which suggests that our model performs well in practice. The uncertainty associated with the estimates of the functional coefficients is also assessed by constructing confidence regions for each estimate. The goodness of fit indices that are commonly used to evaluate the fit of SEMs are developed for the FSEMs introduced in this paper. Overall, the proposed method is a promising approach for modeling functional data in SEMs with functional latent variables.
\end{abstract}


\section{Introduction}\label{sec1}

Structural Equation Models (SEMs) have been successfully used in many fields of science to model how the different aspects of a phenomenon are thought to causally connect to one another.
In SEMs including latent variables, both the observed and the latent variables might be curves over a domain. Several studies in recent years have addressed this kind of situation, for SEMs where all variables are observed, by adapting this setting to the functional data analysis (FDA) context \citep{silverman2002,ramsay2005}. In a functional SEM (FSEM) setting, the directional relationships among the variables are estimated through functional regression models \citep{ramsay2005,wang2016functional}. Examples of this approach are \cite{lindquist2012} and \cite{zhao2018}, who developed a class of functional mediation models, while \cite{zeng2021} and \cite{coffman2022} extended the functional mediation model to deal with sparse and irregular longitudinal data, and with binary-valued outcomes, respectively. A functional mediation model for describing nonlinear relations among the observed variables was studied by \cite{lee2022}.

In some modeling situations where FSEMs would be suited, the variables of interest cannot be directly measured, requiring their estimation through corresponding indicators in a measurement model. While the measurement model itself may be of interest, the primary focus typically lies in the structural component of the model. This structural part specifies the directional relationships among the factors, or between latent factors and observed variables. 
All aforementioned studies focusing on FSEMs primarily address the scenario where all random variables are observed. Despite the many potential applications for FSEMs incorporating latent functional variables, this topic, which explores latent functional variables within FSEMs, has received no attention in the literature so far.

In this work, we focus on FSEMs with latent variables and corresponding indicators sparsely observed over the domain. Handling latent variables in the context of FDA settings can be approached in several ways, and among methods focusing on this issue we find \cite{guo2002}, \cite{morris2006wavelet}, \cite{montagna2012}, \cite{zhu2011robust}, \cite{noh2020}.
We tackle this problem by considering each indicator as a Gaussian random Process (GP) and provide a likelihood-based framework for developing and performing inference in a new class of FSEMs that incorporate latent factors.

The FSEMs developed in this paper are built upon the function-on-function and function-on-scalar regression models. The latent variables in our framework are represented as latent random GPs that are identified through one or more measurement models. Three types of effects for factor loadings are considered in this paper: fixed, concurrent, and historical \citep{malfait2003historical,wang2016functional}.
The fixed effect represents the association between the indicator and the latent factor using a scalar parameter. The concurrent effect describes the pointwise association between the indicator and the latent factor over time, modeled by a smooth function. Historical effects account for the influence of past values of the latent factor on the indicator at a specific point in time, represented by an integral. Our framework supports both scalar and functional covariates, encompassing four types of effects: concurrent, historical, linear, and smooth. Unlike the linear effect, the smooth effect captures the nonlinear influences of scalar covariates.

 To estimate the kernel functions, we employ functional principal component analysis \citep{ramsay2005,rice1991estimating,yao2005functional}.
Since we aim at obtaining smooth estimates for the functional regression coefficients, we employ a penalized likelihood approach in which all smoothing parameters are chosen using a cross-validation method. 


To demonstrate the use of the proposed model, we analyse a dataset from the Health and Retirement Study (HRS) \citep{HRS2024}. The HRS is a nationally representative longitudinal survey conducted from the University of Michigan (US), including more than 12K adults aged 50 and over, and encompassing a comprehensive array of questions concerning health, psychological variables, and income. Initiated in 1992, the HRS is an ongoing study, with participants being interviewed biennially since the study started. This dataset has previously been investigated using longitudinal methods \citep[see][]{newsom2015longitudinal}. 
 In this work, the General Factor of Personality (GFP) is treated as a latent factor identified by the Big Five personality traits. 
 GFP is discussed in the literature and is criticized by different researchers, while others suggest that it could reflect Social Effectiveness (SEff),  influencing behaviors that are socially desirable and advantageous \citep{van2016general}. 
 We employ the proposed FSEM to investigate GFP as a latent factor, and to study the change of the factor loadings across time while estimating the relationship between GFP and different covariates (including `sex' and `having been diagnosed with any type of cancer') in a functional regression framework. 
 

The rest of this paper is organized as follows. In Section \ref{sec:model}, we introduce the measurement model and the structural model for the FSEM, and their finite-dimensional truncated models are given by means of basis expansions.  Section \ref{sec:EM} presents the inferential framework for both estimation and smoothing, which is based on the Expectation-Maximization (EM) algorithm. In Section \ref{conBan},  we address the uncertainty of the estimates for the regression coefficients. Section \ref{gofi} provides extensions for the goodness of fit indices used for the conventional SEM. The performance of the developed model is assessed in Section \ref{sec:simulation} through simulation studies. The proposed model is fitted to real data from the HRS to demonstrate its usefulness in an application context in Section \ref{sec:casestudy}. Finally, the study conclusions and a discussion are provided in Section \ref{sec:conclusion}.

\section{Functional structural equation  model (FSEM)}\label{sec:model}
The core idea of this paper is to handle the problem of functional structural equation modeling with latent variables, in the assumption that both the observed and latent variables take values in a separable Hilbert space $L^2(\tau)$, where $\tau$ is a compact set in $\mathbb{R}$ (e.g., a time interval). 
In this section, the FSEM will be detailed in full, starting with the necessary notation in section \ref{subsec:notation}, followed by the measurement model and the structural part of the FSEM in section \ref{subsec:FSEM}, and their truncated versions in section \ref{subsec:truncated}.

\subsection{Notation}\label{subsec:notation}

Throughout the paper, all the univariate and functional random variables will be represented by regular lowercase letters (e.g., $z$), while vectors are denoted by lowercase bold letters (e.g., $\mathbf{z}$), and matrices are represented by uppercase bold letters (e.g., $\mathbf{Z}$). The difference between the variables and their realizations shall be understood from the context.

We denote via $\{e_r\}_{r\geq 1}$ a set of arbitrary basis functions for the  space $L^2(\tau)$. The  space $L^2(\tau)$ is equipped with the inner product $\langle\cdot,\cdot\rangle$ and the corresponding norm $\Vert\cdot\Vert$. We indicate with $\Vert \cdot\Vert_2$ the usual Euclidean $\ell_2$ norm in $\mathbb{R}^n$. Finally, $\mathbf{1}_{N}$ is a $N$-dimensional unit vector, while $\mathbf{I}_{N}$ is the $N$-dimensional identity matrix. 

\subsection{The functional structural equation  model (FSEM) in infinite dimensions}\label{subsec:FSEM}

Suppose that the latent factors $\eta_1,\eta_2,\ldots,\eta_{q}\in L^2(\tau)$ are measured through variables $y_1,y_2,\ldots,y_p\in L^2(\tau)$, which are in turn observed concatenated with time-dependent measurement errors $\epsilon_{1t},\epsilon_{2t},\ldots,\epsilon_{pt}$, respectively. Assuming to observe $N$ i.i.d. subjects, the functional factor model (FM) for the $j-$th variable in the $i-$th subject $y_{ij}$, $i=1,\ldots,N$ and $j=1,\ldots,p,$ coupled with the measurement error process $\epsilon_{jt},$  can be written as
\begin{align}\label{eq2.3}
y_{ij}(t)&=\beta_{j}(t)+\sum_{m=1}^{q}f_{jm}(\eta_{im},t)a_{jm}+\varepsilon_{ij}(t),\\\label{eq2.3c}
z_{ij}(t)&=y_{ij}(t)+\epsilon_{ijt},
\end{align}
 where $\beta_{j}\in L^2(\tau)$ denote the functional intercepts, and  $\varepsilon_{ij}\in L^2(\tau)$ are the functional unique factors capturing the subject-level functional variation.
 The effect of the latent factor $\eta_{im}$  on the response $y_{ij}$ is flexibly modeled by the term $f_{jm}(\eta_{im},t)$, which can take the form of fixed, concurrent or historical effect. The fixed effect is represented by the term $f_{jm}(\eta_{im},t):=\lambda_{jm}\eta_{im}(t)$, where the factor loading $\lambda_{jm}$ is a scalar parameter. For the concurrent effect, we have $f_{jm}(\eta_{im},t):=\lambda_{jm}(t)\eta_{im}(t)$, while for the historical effect, this term takes the form  $f_{jm}(\eta_{im},t):=\int_{s\leq t}\lambda_{jm}(s,t)\eta_{im}(s)ds$, where the factor loading $\lambda_{jm}(\cdot)$ is instead functional. The term $a_{jm}$ takes the value $1$ if   $y_{ij}$ is regressed on $\eta_{im},$ otherwise is zero. Moreover, the measurement error terms $\epsilon_{ijt}$ are assumed as i.i.d. $\sim N(0,\sigma^2_j)$. Finally,  $\varepsilon_{ij}$ are GPs with $0$ mean and covariance operator $T_j$. 
 We assume each indicator $z_{ij}$  is observed in a sparse or dense subset of the domain at time points $ t_{ij1},t_{ij2},\ldots,t_{ijM_{ij}}\in \tau$.
 Notice that the measurement error $\epsilon_{ijt}$ is independent for each $i,j,t$ and is independent of $\varepsilon_{ij}(t)$, which implies identifiability of the model.

For the structural part, we consider the model 
 \begin{align}\label{eq3.1}\nonumber
     \eta_{im}(t)=&\sum_{n=1}^q s_{mn}^\eta(\eta_{in},t)b^{\eta}_{mn}\\
     &+\sum_{l=1}^{Q}s_{ml}^x(x_{il},t)b^{x}_{ml}+\zeta_{im}(t),
 \end{align}
where $x_1,x_2,\ldots,x_{Q}$ are either scalar or functional observed explanatory variables.  The effects of the latent factors and of these explanatory variables on the latent factors are respectively described by the terms $s_{mn}^\eta(\eta_{in},t)$ and $s_{ml}^x(x_{il},t)$. The term $s_{mn}^\eta(\eta_{in},t)$ can be of the forms concurrent or historical, as this term describes the effect on the functional factor of another functional factor. The term $s_{ml}^x(x_{il},t)$ instead describes the effect of the explanatory variables, and it might be smooth over time or linear, i.e., $s_{ml}^x(x_{il},t)=\gamma_{ml}^x(t)x_{il}$, when the covariate $x_{il}$ is a scalar-valued variable, and concurrent or historical, when $x_{il}$ is a curve over time.  If $\eta_{m}$ is regressed on $\eta_{n}$ and $x_l$, the corresponding terms $b^{\eta}_{mn}$ and $b^{x}_{ml}$ take  value  1, otherwise they are 0. The term $b^{\eta}_{mn}$ is always zero for $m=n$ (to exclude the effect of each functional factor on itself). Furthermore,  $\zeta_{im}$ for $m=1,2,\ldots,q$ are Gaussian with mean functions zero and covariance operators $C_{m}$.

Let  $K_{m}^{\zeta}(s,t)$ be the kernel function of the operator $C_{m}$ and consider $y_{ij'}(t),$ i.e., the first indicator corresponding to the functional factor $\eta_{im}$. Model (\ref{eq2.3}) cannot be uniquely identified unless either we set the first factor loading corresponding to each factor equal to 1, i.e., $f_{j'm}(\eta_{im},t)=\eta_{im}(t)$, or we put the constraint $K_{m}^{\zeta}(t,t)=1$ for all $t\in\tau$. In what follows,  we set the first factor loading corresponding to each factor equal to 1. Furthermore, in the current framework, we suppose the models to be recursive, i.e., feedback effects are not allowed among the latent variables.

\subsection{Truncated model specification}\label{subsec:truncated}
  
In the current section, by rewriting $y_{ij}$ and $\eta_{im}$ according to their corresponding basis expansions, truncated versions of the models (\ref{eq2.3}) and (\ref{eq3.1}) are provided. 

For the measurement part of the model, let  $\mathbf{e}=(e_1,e_2,\ldots,e_J)$ be the vector including the first $J$ basis functions of the space $L^2(\tau)$, and suppose that  $\{\phi_{jr}\}_{1 \leq r \leq J}$ and $\{\nu_{jr}\}_{1 \leq r \leq J}$  are the truncated sets of eigen-functions and eigen-values of the covariance operator $T_{j}$, respectively. Using the Karhunen-Lo\`eve expansion  of $\varepsilon_{ij}(t)$,  the truncated version of the model given in (\ref{eq2.3}) can be written as (see Appendix \ref{ap_a})
\begin{align}\label{eq4.3}
\mathbf{y}_{ij}=\boldsymbol{\beta}_{j}+(\mathbf{E}\mathbf{E}^\top)^{-1}\mathbf{E}\sum_{m= 1}^{q}\mathbf{f}_{ijm}^{'}a_{jm}^{'}+(\mathbf{E}\mathbf{E}^\top)^{-1}\mathbf{E}\sum_{m=1}^q\mathbf{f}_{ijm}a_{jm}+\boldsymbol{\varepsilon}_{ij},
\end{align}
where $a_{jm}^{'}+a_{jm}$ takes either  0 or 1, and $\mathbf{E}=\left[\mathbf{e}(t_{1}),\mathbf{e}(t_2)\ldots,\mathbf{e}(t_{M})\right]\in \mathbb{R}^{J \times M}$ denotes the evaluation matrix of the vector of functions $\mathbf{e}$, where $t_1,t_2,\ldots,t_M$ and the number of time points $M$ may differ for each indicator per sample. Moreover,  $\mathbf{y}_{ij}$ and $\boldsymbol{\beta}_{j} \in \mathbb{R}^J$ denote respectively the vectors of basis coefficients of $y_{ij}$ and of the functional  intercept $\beta_{j}$, while $\mathbf{f}_{ijm}=[f_{jm}(\eta_{im},t_k)]_{1\leq k\leq M}$ and $\mathbf{f}'_{ijm}=[f'_{jm}(\eta_{im},t_k)]_{1\leq k\leq M}$. Finally, $\boldsymbol{\varepsilon}_{ij}$ 
is i.i.d. distributed as a zero mean multivariate Gaussian with variance-covariance matrix $\boldsymbol{\Sigma}_j^{\varepsilon}=\boldsymbol{\Phi}_{j}\mathbf{G}_{j}\boldsymbol{\Phi}_{j}^\top$, where $\mathbf{G}_{j}=\text{diag}\left\{\nu_{j(1)},\ldots, \nu_{j(J)}\right\}$ is the diagonal matrix including the eigen-values of $T_{j}$ sorted in descending order, and $\boldsymbol{\Phi}_{j}$ is the matrix whose elements are the basis coefficients of the eigen-functions of $T_{j}$ in the same order as in $\mathbf{G}_j$, i.e., $\left\{ \boldsymbol{\Phi}_{j} \right\}_{rk}= \langle\phi_{j´(k)},e_r\rangle,$ for $r,k=1,\ldots,J$. 

In Appendix \ref{ap_b},  the different forms for $\mathbf{f}_{ijm}^{'}$ and $\mathbf{f}_{ijm}$ depending on the type of effect are provided. When the effect is fixed, we have $\mathbf{f}_{ijm}^{'}=\mathbf{E}^\top\boldsymbol{\eta}_{im}$ and $\mathbf{f}_{ijm}=(\mathbf{I}_M\otimes\boldsymbol{\eta}_{im})^\top\boldsymbol\omega\lambda_{jm}$. For the concurrent effect we have $\mathbf{f}_{ijm}^{'}=\mathbf{E}^\top\boldsymbol{\eta}_{im}$ and $\mathbf{f}_{ijm}=(\mathbf{I}_M\otimes\boldsymbol{\eta}_{im})^\top\boldsymbol{\Omega}_1\boldsymbol{\lambda}_{jm},$
and for the historical effect we can write $\mathbf{f}_{ijm}^{'}=\boldsymbol{\Delta}\boldsymbol{\eta}_{im}$ and  $\mathbf{f}_{ijm}=(\mathbf{I}_M\otimes\boldsymbol{\eta}_{im})^\top \boldsymbol{\Omega}_2\boldsymbol{\lambda}_{jm}$, where 
\begin{align*}
\boldsymbol\omega&=[\mathbf{e}(t_k)]_{1\leq k\leq M},\\
    \boldsymbol{\Omega}_1 & =\big[\mathbf{e}(t_{k})\mathbf{e}^\top(t_{k})\big]_{1\leq k\leq M},\\
    \boldsymbol{\Omega}_2 & =\left[\int_{s\leq t_{k}}\mathbf{e}(s)(\mathbf{e}^\top(t_{k})\otimes\mathbf{e}^\top(s))ds\right]_{1\leq k\leq M},\\
    \boldsymbol{\Delta} & 
    =\left[\int_{s\leq t_{k}}\mathbf{e}^\top(s)ds\right]_{1\leq k\leq M},
\end{align*}
and $\boldsymbol\lambda_{jm}$ is the vector of basis coefficients for the functional parameter $\lambda_{jm}(t)$. Then, we can rewrite the equation (\ref{eq4.3}) as
\begin{align}\label{eq4.3.3}\nonumber
\mathbf{y}_{ij}=&\boldsymbol{\beta}_{j}+(\mathbf{E}\mathbf{E}^\top)^{-1}\mathbf{E}\sum_{m= 1}^{q}\mathbf{f}_{ijm}^{'}a_{jm}^{'}\\
&+(\mathbf{E}\mathbf{E}^\top)^{-1}\mathbf{E}\sum_{m=1}^q(\mathbf{I}_M\otimes\boldsymbol{\eta}_{im})^\top \boldsymbol{\Omega}^*a_{jm}\boldsymbol{\lambda}_{jm}+\boldsymbol{\varepsilon}_{ij},
\end{align}
in which $\boldsymbol\Omega^*$ equals $\boldsymbol\omega$ for the fixed effect, equals $\boldsymbol\Omega_1$ if the effect is concurrent, and otherwise equals $\boldsymbol\Omega_2$ if the effect is historical.
 Finally, by stacking $\boldsymbol\beta_j$ and $\boldsymbol\lambda_{jm}$ for all $m=1,2,\ldots,q$ into the vector $\boldsymbol\lambda_j$, model (\ref{eq4.3.3}) can be rewritten as
\begin{align}\label{eq4.44}
\mathbf{y}_{ij}=(\mathbf{E}\mathbf{E}^\top)^{-1}\mathbf{E}\sum_{m=1}^q\mathbf{f}^{'}_{ijm}a'_{jm}+\mathbf{F}_i\mathbf{A}_j\boldsymbol\lambda_j+\boldsymbol{\varepsilon}_{ij},
\end{align}
where the block matrix $\mathbf{F}_i$ is defined as $$\mathbf{F}_i=\big[\mathbf{I}_J \vert (\mathbf{E}\mathbf{E}^\top)^{-1}\mathbf{E}(\mathbf{I}_M\otimes\boldsymbol{\eta}_{i1})^\top \boldsymbol{\Omega}^*\vert\cdots\vert (\mathbf{E}\mathbf{E}^\top)^{-1}\mathbf{E}(\mathbf{I}_M\otimes\boldsymbol{\eta}_{iq})^\top \boldsymbol{\Omega}^*\big].$$ Furthermore,  $\mathbf{A}_j$  is defined as 
$$
\mathbf{A}_j= \left[
\begin{array}{cccc}
    \mathbf{I}_J & 0 & \cdots & 0 \\
     0 & a_{j1}\mathbf{I}_{\text{col}(\boldsymbol{\Omega}^*)}
       & \cdots & 0 \\
       \vdots & \vdots & \ddots & \vdots \\
       0 & 0 & \cdots & a_{jq}\mathbf{I}_{\text{col}(\boldsymbol{\Omega}^*)}
\end{array}
\right].
$$
Finally, by stacking the variables $z_{ij}(t_1),z_{ij}(t_2),\ldots,z_{ij}(t_M)$ into the vector $\mathbf{z}_{ij}$ and from (\ref{eq2.3c}), we can write
\begin{equation}\label{eq4.333}
    \mathbf{z}_{ij}=\mathbf{E}^\top\mathbf{y}_{ij}+\boldsymbol{\epsilon}_{ij},
\end{equation}
 where $\boldsymbol{\epsilon}_{ij}$ has Gaussian distribution with mean vector zero and variance-covariance matrix $\sigma^2_j\textbf{I}_{M}$ .

For the structural part of the model, let   $\{\psi_{mr}\}_{1 \leq r \leq J}$ and $\{\mu_{mr}\}_{1\leq r \leq J}$  be the truncated sets of eigen-functions and eigen-values of the covariance operator $C_{m}$. In a similar way to the FM, for the structural model (\ref{eq3.1}) we can write   
\begin{align}\label{eq3.111}
\boldsymbol\eta_{im}=(\mathbf{E}\mathbf{E}^\top)^{-1}\mathbf{E}\left(\sum_{n=1}^q\mathbf{s}_{imn}^\eta b^{\eta}_{mn}
     +\sum_{l=1}^{Q}\mathbf{s}_{iml}^x b^{x}_{ml}\right)+\boldsymbol{\zeta}_{im},
 \end{align}
 where $\mathbf{s}_{imn}^\eta=[ s_{mn}^\eta(\eta_{in},t_k)]_{1\leq k\leq M}$, $\mathbf{s}_{iml}^x=[ s_{ml}^x(x_{il},t_k)]_{1\leq k\leq M}$, and $\boldsymbol{\eta}_{im}$ is the vector of basis coefficients of $\eta_{im}$. 
 
 Moreover, 
 $\boldsymbol{\zeta}_{im}$ has Gaussian distribution with mean zero and covariance matrix $\boldsymbol{\Sigma}_m^{\zeta}=\boldsymbol{\Psi}_{m}\mathbf{H}_{m}\boldsymbol{\Psi}_{m}^\top$, in which 
$\mathbf{H}_{m}=\text{diag}\left\{\mu_{m(1)},\ldots, \mu_{m(J)}\right\}$ is the diagonal matrix including the eigen-values of $C_{m}$ sorted in descending order, and $\boldsymbol{\Psi}_{m}$ is the matrix whose elements are the basis coefficients of the eigen-functions of $C_{m}$ in the same order as in $\mathbf{H}_m$, i.e., $\left\{ \boldsymbol{\Psi}_{m} \right\}_{rk}= \langle\psi_{m(k)},e_r\rangle,$ for $r,k=1,\ldots,J$.

Similarly to what we have seen for the measurement model,  $\mathbf{s}_{imn}^\eta$ and $\mathbf{s}_{iml}^x$ can be written in several forms.
In the case of concurrent effects, we can write $\mathbf{s}_{imn}^\eta=(\mathbf{I}_M\otimes\boldsymbol{\eta}_{in})^\top\boldsymbol{\Omega}_1\boldsymbol{\gamma}_{mn}^\eta$ and $\mathbf{s}_{iml}^x=(\mathbf{I}_M\otimes\mathbf{x}_{il})^\top\boldsymbol{\Omega}_1\boldsymbol{\gamma}_{ml}^x$. For the historical effects we have $\mathbf{s}_{imn}^\eta=(\mathbf{I}_M\otimes\boldsymbol{\eta}_{in})^\top \boldsymbol{\Omega}_2\boldsymbol{\gamma}_{mn}^\eta$ and $\mathbf{s}_{iml}^x=(\mathbf{I}_M\otimes\mathbf{x}_{il})^\top \boldsymbol{\Omega}_2\boldsymbol{\gamma}_{ml}^x$, where $\boldsymbol{\gamma}^\eta_{mn}$ and $\boldsymbol{\gamma}^x_{ml}$ denote the vectors of basis coefficients for the functional parameters $\gamma^\eta_{mn}(t)$ and $\gamma^x_{ml}(t)$, respectively. Finally, a scalar covariate (corresponding to a smooth or linear effect) is denoted via $x_{il}$, while for  curve covariates $x_{il}(t)$ with concurrent or historical effects, $\mathbf{x}_{il}$ indicates the vector of basis coefficients for the functional covariate.

Let $\{h_{lr}\}_{r\geq 1}$ be another set of arbitrary basis functions for the Hilbert space $L^2(\tau_{l})$, where $\tau_{l}$ denotes the range of the covariate $x_l$ for $l=1,2,\ldots,Q$. For the smooth and linear effects of the covariates in the structural model given by (\ref{eq3.111}), we obtain respectively  $\mathbf{s}_{iml}^x=\big(\mathbf{E}\otimes\mathbf{h}_l(x_{il})\big)^\top\boldsymbol\gamma_{ml}^x$ and $\mathbf{s}_{iml}^x=\mathbf{E}^\top\boldsymbol\gamma_{ml}^x x_{il}$, in which  $\mathbf{h}_{l}(x_{il})=\big(h_{lr}(x_{il})\big)_{1\leq r\leq J}$ and $\boldsymbol{\gamma}_{ml}^x$ denote the vectors of unknown parameters (see Appendix \ref{ap_c}).

Consider the matrix $$\mathbf{S}_i^\eta=\big[ (\mathbf{E}\mathbf{E}^\top)^{-1}\mathbf{E}(\mathbf{I}_M\otimes\boldsymbol{\eta}_{i1})^\top \boldsymbol{\Omega}^*\vert\cdots\vert (\mathbf{E}\mathbf{E}^\top)^{-1}\mathbf{E}(\mathbf{I}_M\otimes\boldsymbol{\eta}_{iq})^\top \boldsymbol{\Omega}^*\big].$$
Similarly to the FM case, model (\ref{eq3.111}) can be rewritten as
\begin{align}\label{eq4.444}
\boldsymbol\eta_{im}=\mathbf{S}_i\mathbf{B}_{m}\boldsymbol\gamma_{m}+\boldsymbol{\zeta}_{im},
\end{align}
 where $\boldsymbol\gamma_m$ is the vector of basis coefficients for all functional parameters, and $\mathbf{S}_i=[\mathbf{S}_i^\eta\vert(\mathbf{E}\mathbf{E}^\top)^{-1}\mathbf{E}\mathbf{S}_{i1}^x\vert\cdots\vert(\mathbf{E}\mathbf{E}^\top)^{-1}\mathbf{E}\mathbf{S}_{iQ}^x]$, where $\mathbf{S}_{il}^x$ for $l=1,2,\ldots,Q$ can be in different forms depending on the type of effect of $x_l$ on $\eta_m$. Moreover,  $\mathbf{B}_{m}=\text{diag}\left\{\mathbf{B}_m^{\eta},\mathbf{B}_m^x\right\}$, in which
 $$
\mathbf{B}_m^\eta= \left[
\begin{array}{cccc}
    b_{m1}^\eta\mathbf{I}_{\text{col}(\boldsymbol{\Omega}^*)} & 0 & \cdots & 0 \\
     0 & b_{m2}^\eta\mathbf{I}_{\text{col}(\boldsymbol{\Omega}^*)}
       & \cdots & 0 \\
       \vdots & \vdots & \ddots & \vdots \\
       0 & 0 & \cdots & b_{mq}^\eta\mathbf{I}_{\text{col}(\boldsymbol{\Omega}^*)}
\end{array}
\right],
$$
 and
 $$
\mathbf{B}_m^x= \left[
\begin{array}{cccc}
    b_{m1}^x\mathbf{I}_{\text{col}(\mathbf{S}_{1m1}^x)} & 0 & \cdots & 0 \\
     0 & b_{m2}^x\mathbf{I}_{\text{col}(\mathbf{S}_{1m2}^x)}
       & \cdots & 0 \\
       \vdots & \vdots & \ddots & \vdots \\
       0 & 0 & \cdots & b_{mQ}^x\mathbf{I}_{\text{col}(\mathbf{S}_{1mQ}^x)}
\end{array}\right].
$$

\section{Estimation procedure}\label{sec:EM}
In this section, the Expectation-Maximization (EM) algorithm employed to provide the maximum likelihood estimates of both scalar and functional parameters in the model is described in full detail. In order to estimate the functional parameters,  their corresponding basis coefficients are estimated and, subsequently, the parameter curves are estimated using the basis expansions.
In what follows, to provide estimates of the parameters in the measurement and structural models (\ref{eq2.3c}) and (\ref{eq3.1}), we consider the following assumptions:
\setlength{\leftmargini}{17pt}
\begin{enumerate}[label=(\alph*)]
\item The measurement model is reflective, i.e., the indicators are explained by one or more common factors, and the structural models are recursive, meaning that there are no permitted feedback effects between the latent variables;
\item the  $y_j$'s    are conditionally independent given the latent variables for $j=1,2,\ldots,p$;
     \item the measurement error  terms $\epsilon_{ijt}$'s are supposed to be independent  for $j=1,2,\ldots,p$ and across time;
    \item the terms $\epsilon_{ijt}$'s are independent of $\varepsilon_{ij}$'s;
     \item the latent variables  $\eta_m$'s for  $m=1,2,\ldots,q$, are conditionally independent given the observed and latent variables.
\end{enumerate}

\subsection{EM algorithm implementing the smoothing procedure}\label{sec3.3}
To write the complete-data log-likelihood function, we stack $\mathbf{y}_{ij}$, $\mathbf{z}_{ij}$ and $\boldsymbol\eta_{im}$ first for all subjects and then for  $j=1,2,\ldots,p$ and $m=1,2,\ldots,q$ into the vectors $\mathbf{y}$, $\mathbf{z}$ and $\boldsymbol\eta$, respectively.  We also suppose $\mathbf{x}$ is the vector of all explanatory variables in the structural model, stacked for all subjects. Then, setting $\boldsymbol\theta$ as the set of all model parameters and
based on the assumptions considered for the functional and structural parts of the FSEM,   
the complete-data log-likelihood function  is written as $\ell_{\text{com}}(\boldsymbol{\theta}\vert\mathbf{z},\mathbf{y},\boldsymbol{\eta},\mathbf{x})=\log f(\mathbf{z},\mathbf{y},\boldsymbol{\eta},\mathbf{x};\boldsymbol{\theta})$,
where $f(\mathbf{z},\mathbf{y},\boldsymbol{\eta},\mathbf{x};\boldsymbol{\theta})$ denotes the joint probability density function (pdf) of $\mathbf{z},\mathbf{y},\boldsymbol{\eta},\mathbf{x}$ and the set of model parameters $\boldsymbol{\theta}$.

In the E-step at iteration $m'+1$, we compute the expectation of the complete-data log-likelihood function with respect to the joint distribution of the latent variables $\mathbf{y}$ and $\boldsymbol{\eta}$  given the observed variables $\mathbf{z}$ and $\mathbf{x}$ and the model parameters at iteration $m'$ as
 $$Q(\boldsymbol{\theta}\mid \boldsymbol{\theta}^{(m')})=\mathbb{E}_{\mathbf{y},\boldsymbol{\eta}\vert \mathbf{z},\mathbf{x};\boldsymbol{\theta}^{(m')}}[\ell_{\text{com}}(\boldsymbol{\theta}\vert\mathbf{z},\mathbf{y},\boldsymbol{\eta},\mathbf{x})].$$

We implement the smoothing procedure in the M-step to extract smooth estimates for the functional coefficients. For this purpose, let $\boldsymbol\alpha$ be the vector stacking all smoothing parameters corresponding to functional coefficients from the measurement and structural part of the FSEM, respectively, i.e. $\alpha_j$ and $\alpha_m$ for $j=1,2,\ldots,p$ and $m=1,2,\ldots,q$. 
 In the M-step, we maximize the expectation  
\begin{align*}
Q_{\boldsymbol{\alpha}}(\boldsymbol{\theta}\mid \boldsymbol{\theta}^{(m')})=&\mathbb{E}_{\mathbf{y}\vert \mathbf{z},\mathbf{x};\boldsymbol{\theta}^{(m')}}\mathbb{E}_{\boldsymbol{\eta}\vert \mathbf{y},\mathbf{x};\boldsymbol{\theta}^{(m')}}[\log f( \mathbf{z},\mathbf{y},\boldsymbol\eta, \mathbf{x};\boldsymbol{\theta})] \\
=&\mathbb{E}_{\mathbf{y}\vert \mathbf{z},\mathbf{x};\boldsymbol{\theta}^{(m')}}[\log f( \mathbf{z}\vert \mathbf{y},\mathbf{x};\boldsymbol{\theta})]\\
&+\mathbb{E}_{\mathbf{y}\vert \mathbf{z},\mathbf{x};\boldsymbol{\theta}^{(m')}}\mathbb{E}_{\boldsymbol{\eta}\vert \mathbf{y},\mathbf{x};\boldsymbol{\theta}^{(m')}}[\log f( \mathbf{y}\vert \boldsymbol{\eta},\mathbf{x};\boldsymbol{\theta})]\\
&+\mathbb{E}_{\mathbf{y}\vert \mathbf{z},\mathbf{x};\boldsymbol{\theta}^{(m')}}\mathbb{E}_{\boldsymbol{\eta}\vert \mathbf{y},\mathbf{x};\boldsymbol{\theta}^{(m')}}[\log f( \boldsymbol{\eta}\vert \mathbf{x};\boldsymbol{\theta})]\\
&-\alpha_j^{\lambda}\sum_{j=1}^p \text{pen}(\boldsymbol\lambda_j)-\alpha_m^{\gamma}\sum_{m=1}^q \text{pen}(\boldsymbol\gamma_m), 
\end{align*}
where $\text{pen}(\boldsymbol\lambda_j)=\boldsymbol\lambda_j^\top \mathbf{P}^\lambda_j\boldsymbol\lambda_j$ and $\text{pen}(\boldsymbol\gamma_m)=\boldsymbol\gamma_m^\top \mathbf{P}^\gamma_m\boldsymbol\gamma_m$ are the penalty terms corresponding to the parameters $\boldsymbol\lambda_j$ and $\boldsymbol\gamma_m$, respectively. Further, the penalty matrices $\mathbf{P}_j^{\lambda}$ and $\mathbf{P}_m^{\gamma}$ are defined as
$$
\mathbf{P}^\lambda_j= \left[
\begin{array}{cccc}
    \mathbf{P}_{j1}^\lambda & 0 & \cdots & 0 \\
     0 & \mathbf{P}_{j2}^\lambda
       & \cdots & 0 \\
       \vdots & \vdots & \ddots & \vdots \\
       0 & 0 & \cdots & \mathbf{P}_{jq}^\lambda
\end{array}
\right],
$$
and $\mathbf{P}^\gamma_m=\text{diag}\{\mathbf{P}^{\gamma^\eta}_m,\mathbf{P}^{\gamma^x}_m\}$, in which
$$
\mathbf{P}^{\gamma^\eta}_m= \left[
\begin{array}{cccc}
    \mathbf{P}^{\gamma^\eta}_{m1} & 0 & \cdots & 0 \\
     0 & \mathbf{P}^{\gamma^\eta}_{m2}
       & \cdots & 0 \\
       \vdots & \vdots & \ddots & \vdots \\
       0 & 0 & \cdots & \mathbf{P}^{\gamma^\eta}_{mq}
\end{array}
\right],
$$ 
and
$$
\mathbf{P}^{\gamma^x}_m= \left[
\begin{array}{cccc}
    \mathbf{P}^{\gamma^x}_{m1} & 0 & \cdots & 0 \\
     0 & \mathbf{P}^{\gamma^x}_{m2}
       & \cdots & 0 \\
       \vdots & \vdots & \ddots & \vdots \\
       0 & 0 & \cdots & \mathbf{P}^{\gamma^x}_{mQ}
\end{array}
\right].
$$
 Let $\mathbf{e}^{(l)}(t)=[e_{r}^{(l)}(t)]_{1\leq r\leq J}$ be the vector of the $l^
{\text{th}}$ order derivatives of $e_{r}(t)$. Depending on the type of effects in the measurement model (\ref{eq2.3})  and structural model (\ref{eq3.1}), $\mathbf{P}^\lambda_{jm}$, $\mathbf{P}_{mn}^{\gamma^\eta}$ and $\mathbf{P}_{mr}^{\gamma^x}$   can take different forms, and specifically $\int_{\tau} \mathbf{e}^{(2)}(t)\mathbf{e}^{(2)\top}(t) dt$ for the concurrent effect, and 
\begin{align*}
&\int_\tau\int_\tau \left(\mathbf{e}^{(1)}(t)\otimes\mathbf{e}^{(1)}(s)\right)\left(\mathbf{e}^{(1)\top}(t)\otimes\mathbf{e}^{(1)\top}(s)\right)dsdt\\
&+\int_\tau\int_\tau \left(\mathbf{e}(t)\otimes\mathbf{e}^{(2)}(s)\right)\left(\mathbf{e}^{\top}(t)\otimes\mathbf{e}^{(2)\top}(s)\right)dsdt\\
&+\int_\tau\int_\tau \left(\mathbf{e}^{(2)}(t)\otimes\mathbf{e}(s)\right)\left(\mathbf{e}^{(2)\top}(t)\otimes\mathbf{e}^{\top}(s)\right)dsdt,
\end{align*}
for the historical effect. Moreover,   $$\mathbf{P}_{mr}^{\gamma^x} = \int_\tau \big(\mathbf{e}^{(2)}(t)\otimes\mathbf{h}_r(x_r)\big)\big(\mathbf{e}^{(2)\top}(t)\otimes\mathbf{h}_r^\top(x_r)\big)dt,
$$
 when the effect of $\eta_m$ on $x_r$ is smooth. 

Let us stack  the matrices $\mathbf{F}_i$ and $\mathbf{S}_i$ for $i=1,2,\ldots,N$ respectively into the block matrices  $\mathbf{F}$ and $\mathbf{S}$ and the vectors $(\mathbf{E}\mathbf{E}^\top)^{-1}\mathbf{E}\mathbf{f}_{ijm}^{'}$, $\mathbf{y}_{ij}$ and $\boldsymbol{\eta}_{im}$ for $i=1,2,\ldots,N$ into the vectors $\mathbf{f}_{jm}^{'}$, $\mathbf{y}_j$ and $\boldsymbol{\eta}_m$, respectively.
The parameter estimates at iteration $m'+1$ are updated as
\begin{align*}
\boldsymbol\lambda_j^{(m'+1)}=&\big(\mathbf{A}_j\mathbb{E}_{\mathbf{y}\vert\mathbf{z},\mathbf{x};\boldsymbol{\theta}^{(m')}}\mathbb{E}_{\boldsymbol{\eta}\vert \mathbf{y},\mathbf{x};\boldsymbol{\theta}^{(m')}}\big[\mathbf{F}^\top(\mathbf{I}_{N}\otimes\boldsymbol\Sigma_j^{\varepsilon^{(m'+1)}})^{-1}\mathbf{F}\big]\mathbf{A}_j+\alpha_j^\lambda\mathbf{P}_j^\lambda\big)^{-1}\\
&\times\mathbf{A}_j\mathbb{E}_{\mathbf{y}\vert\mathbf{z},\mathbf{x};\boldsymbol{\theta}^{(m')}}\mathbb{E}_{\boldsymbol{\eta}\vert \mathbf{y},\mathbf{x};\boldsymbol{\theta}^{(m')}}\big[\mathbf{F}^\top(\mathbf{I}_{N}\otimes\boldsymbol\Sigma_j^{\varepsilon^{(m'+1)}})^{-1}
\big(\mathbf{y}_{j}-\sum_{m=1}^q\mathbf{f}_{jm}^{'}a_{jm}^{'}\big)\big],\\
\boldsymbol\Sigma_j^{\varepsilon^{(m'+1)}}=&\frac{1}{N}\sum_{i=1}^{N}
\mathbb{E}_{\mathbf{y}\vert \mathbf{z},\mathbf{x};\boldsymbol{\theta}^{(m')}}\mathbb{E}_{\boldsymbol{\eta}\vert \mathbf{y},\mathbf{x};\boldsymbol{\theta}^{(m')}}\big[\big(\mathbf{y}_{ij}-(\mathbf{E}\mathbf{E}^\top)^{-1}\mathbf{E}\sum_{m=1}^q\mathbf{f}_{jm}^{'}a_{jm}^{'}-\mathbf{F}_{i}\mathbf{A}_j\boldsymbol\lambda_j^{(m'+1)}\big)\\
&\times\big(\mathbf{y}_{ij}-(\mathbf{E}\mathbf{E}^\top)^{-1}\mathbf{E}\sum_{m=1}^q\mathbf{f}_{jm}^{'}a_{jm}^{'}-\mathbf{F}_{i}\mathbf{A}_j\boldsymbol\lambda_j^{(m'+1)}\big)^\top\big],\\
\sigma^{2(m'+1)}_j=&\frac{1}{(\sum_{i=1}^{N}M_{ij})}\sum_{i=1}^{N}\mathbb{E}_{\mathbf{y}\vert \mathbf{z},\mathbf{x};\boldsymbol{\theta}^{(m')}}\big(\mathbf{z}_{ij}-\mathbf{E}_{ij}^\top\mathbf{y}_{ij}\big)^\top\big(\mathbf{z}_{ij}-\mathbf{E}_{ij}^\top\mathbf{y}_{ij}\big),
\end{align*}
for $j=1,2,\ldots,p$, and
\begin{align*}
\boldsymbol\gamma_{m}^{(m'+1)}=&\big(\mathbf{B}_{m}\mathbb{E}_{\mathbf{y}\vert \mathbf{z},\mathbf{x};\boldsymbol{\theta}^{(m')}}\mathbb{E}_{\boldsymbol{\eta}\vert \mathbf{y},\mathbf{x};\boldsymbol{\theta}^{(m')}}\big[\mathbf{S}^\top(\mathbf{I}_{N}\otimes\boldsymbol\Sigma_{m}^{\zeta^{(m'+1)}})^{-1}\mathbf{S}\big]\mathbf{B}_{m}+\alpha_m^\gamma\mathbf{P}_m^\gamma\big)^{-1}\\
&\times\mathbf{B}_{m}\mathbb{E}_{\mathbf{y}\vert \mathbf{z},\mathbf{x};\boldsymbol{\theta}^{(m')}}\mathbb{E}_{\boldsymbol{\eta}\vert \mathbf{y},\mathbf{x};\boldsymbol{\theta}^{(m')}}\big[\mathbf{S}^\top(\mathbf{I}_{N}\otimes\boldsymbol\Sigma_{m}^{\zeta^{(m'+1)}})^{-1}\boldsymbol\eta_{m}\big],\\
\boldsymbol\Sigma_{m}^{\zeta^{(m'+1)}}=&\frac{1}{N}\sum_{i=1}^{N}
\mathbb{E}_{\mathbf{y}\vert \mathbf{z},\mathbf{x};\boldsymbol{\theta}^{(m')}}\mathbb{E}_{\boldsymbol{\eta}\vert \mathbf{y},\mathbf{x};\boldsymbol{\theta}^{(m')}}\big[\big(\boldsymbol\eta_{im}-\mathbf{S}_{i}\mathbf{B}_{m}\boldsymbol\gamma_{m}^{(m'+1)}\big)\\
&\times\big(\boldsymbol\eta_{im}-\mathbf{S}_{i}\mathbf{B}_{m}\boldsymbol\gamma_{m}^{(m'+1)}\big)^\top\big],
\end{align*}
for $m=1,2,\ldots,q$.

We evaluate the convergence of the EM algorithm by looking at the two following criteria
\begin{align*}
\text{tol}_{\sigma^2} = & \max\big(\Vert\sigma^{2^{(m'+1)}}_j-\sigma^{2^{(m')}}_j\Vert_2\big)_{j=1,2,\ldots,p} \\
\text{tol}_{coef}  = & \max\left\{\Vert \boldsymbol{\lambda}_{jm}^{-(m'+1)}-\boldsymbol{\lambda}_{jm}^{-(m')}\Vert_2, \Vert \boldsymbol{\beta}_{j}^{(m'+1)}-\boldsymbol{\beta}_{j}^{(m')}\Vert_2, \right. \\
& \quad \quad \left. \Vert \boldsymbol{\gamma}_{mn}^{(m'+1)}-\boldsymbol{\gamma}_{mn}^{(m')}\Vert_2, \Vert \boldsymbol{\gamma}_{ml}^{(m'+1)}-\boldsymbol{\gamma}_{ml}^{(m')}\Vert_2
\right\},
\end{align*}
where the maximum in $tol_{coef}$ is taken over $j=1,2,\ldots,p;$ $m,n=1,2,\ldots,q;$ $l=1,2,\ldots,Q$, and where $\boldsymbol{\lambda}_{jm}^{-(m')}$ denotes the vector $\boldsymbol{\lambda}_{jm}$ at iteration $m'$ excluding the estimates for the intercept $\beta_j$.
 To evaluate the convergence of the EM, we compare the values of $tol_{\sigma^2}$ and $tol_{coef}$ with a pre-specified tolerance, which can be different for the two criteria.  

\subsection{Generator distributions}\label{s.gen}

The updates of parameter estimates given by the expectations described in Section \ref{sec3.3} can be computed numerically by applying the Monte Carlo approach.

Suppose we stack $\boldsymbol{\eta}_{im}$ for $m=1,2,\ldots,q$ into $\boldsymbol{\eta}_i$. If we define the  matrices $
\mathbf{A}_j^{'}=\text{diag}(a_{jm}^{'}\mathbf{I}_{J})_{1\leq m\leq q}$  and $
\mathbf{A}_j^*=\text{diag}\left(a_{jm}\mathbf{I}_{J}\right)_{1\leq m\leq q}$, as well as the matrices $\mathbf{B}_{m}^{\eta^*}=\text{diag}(b_{mn}^\eta\mathbf{I}_{J})_{1\leq n\leq q}$ and $\mathbf{B}_{m}^{x^*}=\text{diag}(b_{ml}^x\mathbf{I}_{col(\boldsymbol\Gamma^x_{ml})})_{1\leq l\leq Q}$, the models (\ref{eq4.44}) and (\ref{eq4.444}) can be rewritten as 
\begin{align}\label{eq.sgen1}
\mathbf{y}_{ij}=\boldsymbol\beta_{j}+\boldsymbol\Lambda_{j}^{'}\mathbf{A}_j^{'}\boldsymbol\eta_i+\boldsymbol\Lambda_{j}\mathbf{A}_j^*\boldsymbol\eta_i+\boldsymbol\varepsilon_{ij},
\end{align}
and 
\begin{align}\label{eq.sgen2}
\boldsymbol\eta_{im}=\boldsymbol\Gamma_{m}^\eta\mathbf{B}_{m}^{\eta^*}\boldsymbol\eta_i+\boldsymbol\Gamma_{m}^x\mathbf{B}_{m}^{x^*}\mathbf{x}_i+\boldsymbol\zeta_{im},
\end{align}
where $\boldsymbol\Lambda_j=(\mathbf{E}\mathbf{E}^\top)^{-1}\mathbf{E}[\boldsymbol{\Lambda}_{j1}\vert\boldsymbol{\Lambda}_{j2}\vert\ldots\vert\boldsymbol{\Lambda}_{jq}]$, and the same definition holds for $\boldsymbol\Lambda_j^{'}$ with $\boldsymbol{\Lambda}_{jm}^{'}$ for $m=1,\ldots,q$ instead. When the effect of $\eta_m$ on $y_j$ is fixed we have $\boldsymbol{\Lambda}_{jm}^{'}=\mathbf{E}^\top$ and $\boldsymbol{\Lambda}_{jm}=\lambda_{jm}\mathbf{E}^\top$, while for the  concurrent effect we obtain  $\boldsymbol{\Lambda}_{jm}^{'}=\mathbf{E}^\top$ and $\boldsymbol{\Lambda}_{jm}=(\mathbf{I}_M\otimes\boldsymbol\lambda_{jm}^\top)\boldsymbol\Omega_1.$ When the effect is instead historical, we have $\boldsymbol{\Lambda}_{jm}^{'}=\boldsymbol{\Delta}$ and $\boldsymbol{\Lambda}_{jm}=(\mathbf{I}_M\otimes\boldsymbol\lambda_{jm}^\top)\boldsymbol\Omega_2^{*}$ with $\boldsymbol{\Omega}_2^*=\big[\int_{s\leq t_{k}}(\mathbf{e}(t_{k})\otimes\mathbf{e}(s))\mathbf{e}^\top(s)ds\big]_{1\leq k\leq M}$.   Furthermore, $\mathbf{x}_i=(\mathbf{x}_{il})_{1\leq l\leq Q}$ and $\boldsymbol\Gamma_{m}^\eta=(\mathbf{E}\mathbf{E}^\top)^{-1}\mathbf{E}\big[\boldsymbol\Gamma^\eta_{m1}\vert\boldsymbol\Gamma^\eta_{m2}\vert\ldots\vert\boldsymbol\Gamma^\eta_{mq}\big]$, where $\boldsymbol\Gamma^\eta_{mn}=(\mathbf{I}_M\otimes\boldsymbol\gamma_{mn}^{\eta})^\top\boldsymbol\Omega_1$ in the case of concurrent effect of $\eta_{n}$ on $\eta_m$ and $\boldsymbol\Gamma^\eta_{mn}=(\mathbf{I}_M\otimes\boldsymbol\gamma_{mn}^{\eta})^\top\boldsymbol\Omega_2^*$, when historical. Finally, $\boldsymbol\Gamma_{m}^x=(\mathbf{E}\mathbf{E}^\top)^{-1}\mathbf{E}\big[\boldsymbol\Gamma^x_{m1}\vert\boldsymbol\Gamma^x_{m2}\vert\ldots\vert\boldsymbol\Gamma^x_{mQ}\big]$, where $\boldsymbol\Gamma^x_{ml}$ can take the following different forms: $(\mathbf{I}_M\otimes\boldsymbol\gamma_{ml}^{x})^\top\boldsymbol{\Omega}_1$, $(\mathbf{I}_M\otimes\boldsymbol\gamma_{ml}^{x})^\top\boldsymbol{\Omega}_2^*$,  $\mathbf{E}^\top\boldsymbol\Delta_{ml}$, or $\mathbf{E}^\top\boldsymbol\gamma_{ml}^x$ depending on the type of effect of the covariate being respectively concurrent, historical, smooth or linear. In the form corresponding to the smooth effect, $\boldsymbol\Delta_{ml}$ is the matrix of basis coefficients. Depending on the different types of effects,  $\mathbf{x}_{il}$ indicates the vector of basis coefficients of $x_l$ when its effect is either concurrent or historical, $\mathbf{x}_{il}=\mathbf{h}_r(x_l)$ if the effect is smooth, and $\mathbf{x}_{il}=x_{il}$ in case of linear effect.

Model (\ref{eq.sgen1}) can be written as $$\mathbf{y}=\boldsymbol{\beta}+(\boldsymbol\Lambda+\boldsymbol\Lambda')\mathbf{W}_{\boldsymbol\eta}\boldsymbol\eta+\boldsymbol\varepsilon,$$ where 
$\boldsymbol\beta=(\mathbf{1}_{N}\otimes\boldsymbol\beta_j)_{1\leq j\leq p}$, $\boldsymbol\Lambda'=[\mathbf{I}_{N}\otimes(\boldsymbol\Lambda_j^{'}\mathbf{A}_j^{'})]_{1\leq j\leq p}$,  $\boldsymbol\Lambda=[\mathbf{I}_{N}\otimes(\boldsymbol\Lambda_j\mathbf{A}_j^*)]_{1\leq j\leq p}$. $\mathbf{W}_{\boldsymbol\eta}$ is a weight matrix associated to the vector $\boldsymbol\eta,$ and $\boldsymbol\varepsilon$ has Gaussian distribution with mean zero and covariance matrix $\text{diag}\left\{\mathbf{I}_{N}\otimes \boldsymbol\Sigma_j^\varepsilon\right\}_{1\leq j\leq p}$.

For the structural model (\ref{eq.sgen2}) we can write
\begin{equation*}
\boldsymbol\eta=\boldsymbol\Gamma^\eta\mathbf{W}_{\boldsymbol\eta}\boldsymbol\eta+\boldsymbol\Gamma^x\mathbf{x}+\boldsymbol\zeta,
\end{equation*}
where $\boldsymbol\Gamma^\eta=\big[\mathbf{I}_{N}\otimes\big(\boldsymbol\Gamma_{m}^\eta\mathbf{B}_{m}^{\eta^*}\big)\big]_{1\leq m\leq q}$,  $\boldsymbol\Gamma^x=\big[\mathbf{I}_{N}\otimes(\boldsymbol\Gamma_{r}^x\mathbf{B}_{r}^{x^*})\big]_{1\leq r\leq Q}$  and $\boldsymbol\zeta$ is Gaussian with zero mean and  covariance matrix $\text{diag}\{\mathbf{I}_{N}\otimes\boldsymbol\Sigma_{m}^\zeta\}_{1\leq m\leq q}$. 

This implies that $\boldsymbol\eta\sim\mathcal{N}(\boldsymbol\mu_{\boldsymbol\eta}, \boldsymbol\Sigma_{\boldsymbol\eta})$, with $\boldsymbol\mu_{\boldsymbol\eta}=(\mathbf{I}_{qNJ}-\boldsymbol\Gamma^\eta\mathbf{W}_{\boldsymbol\eta})^{-1}\boldsymbol\Gamma^x\mathbf{x}$ and $\boldsymbol\Sigma_{\boldsymbol\eta}=(\mathbf{I}_{qNJ}-\boldsymbol\Gamma^\eta\mathbf{W}_{\boldsymbol\eta})^{-1}\text{diag}\{\mathbf{I}_{N}\otimes\boldsymbol\Sigma_{m}^\zeta\}_{1\leq m\leq q}(\mathbf{I}_{qNJ}-\boldsymbol\Gamma^\eta\mathbf{W}_{\boldsymbol\eta})^{-\top}$. Analogously, $\mathbf{y}\sim\mathcal{N}(\boldsymbol\mu_{\mathbf{y}}, \boldsymbol\Sigma_{\mathbf{y}}),$ with $\boldsymbol\mu_{\mathbf{y}}=\boldsymbol{\beta}+(\boldsymbol\Lambda+\boldsymbol\Lambda')\mathbf{W}_{\boldsymbol\eta}\boldsymbol\mu_{\boldsymbol\eta}$ and $\boldsymbol\Sigma_{\mathbf{y}}=(\boldsymbol\Lambda+\boldsymbol\Lambda')\mathbf{W}_{\boldsymbol\eta}\boldsymbol\Sigma_{\boldsymbol\eta}\mathbf{W}_{\boldsymbol\eta}(\boldsymbol\Lambda+\boldsymbol\Lambda')^\top+\text{diag}\left\{\mathbf{I}_{N}\otimes \boldsymbol\Sigma_j^\varepsilon\right\}_{1\leq j\leq p}$. 

Accordingly, from (\ref{eq4.333}) and defining $\mathbf{E}^*=\text{diag}\{\text{diag}(\mathbf{E}^\top_{ij})_{1\leq i\leq N}\}_{1\leq j\leq p}$, where $\mathbf{E}_{ij}=(\mathbf{e}(t_{ij1}),\mathbf{e}(t_{ij2}),\ldots,\mathbf{e}(t_{ijM_{ij}}))$, then we have $\mathbf{z}\sim\mathcal{N}(\boldsymbol\mu_{\mathbf{z}},\boldsymbol\Sigma_{\mathbf{z}}),$ with $\boldsymbol\mu_{\mathbf{z}}=\mathbf{E}^*\boldsymbol\mu_{\mathbf{y}}$ and $\boldsymbol\Sigma_{\mathbf{z}}=\mathbf{E}^*\boldsymbol\Sigma_{\mathbf{y}}\mathbf{E}^{*\top}+\text{diag}\{\sigma^{2}_j\text{diag}(\mathbf{I}_{M_{ij}})_{1\leq i\leq N}\}_{1\leq j\leq p}$.

Thanks to the normality of $\mathbf{y}$, $\mathbf{z}$ and $\boldsymbol\eta$, the joint distributions of $(\mathbf{y},\mathbf{z})$ and  $(\boldsymbol\eta,\mathbf{y})$  are also Gaussian. Thus, the generator distributions are also Gaussian, and their parameters can be simply derived. 
The selection of the weight matrix $\mathbf{W}_{\boldsymbol\eta}$ is quite crucial and it depends on the specific model, with the flexibility to vary on a case-by-case basis.

\section{Confidence bands}\label{conBan}

In this section, we quantify the uncertainty of the estimates for 
the factor loading parameters in the measurement model (\ref{eq2.3c}) and for the regression coefficients in the structural model (\ref{eq3.1}). For this purpose, the confidence regions are constructed based on the geometric approach developed by \cite{choi2018geometric}. To construct the confidence regions, we need to estimate the covariance matrices of the estimates for $\boldsymbol{\lambda}_j$ and $\boldsymbol{\gamma}_m$ at the final step of the EM algorithm, denoted by $\hat{\boldsymbol{\Sigma}}_{\boldsymbol{\lambda}_j}$ and $\hat{\boldsymbol{\Sigma}}_{\boldsymbol{\gamma}_m}$. We employ a non-parametric bootstrapping method to extract the covariance matrices $\hat{\boldsymbol{\Sigma}}_{\boldsymbol{\lambda}_j}$ and $\hat{\boldsymbol{\Sigma}}_{\boldsymbol{\gamma}_m}$. Considering the matrices $\text{diag}(\hat{\boldsymbol\Sigma}_{\boldsymbol\lambda_j})=\text{diag}(\hat{\boldsymbol\Sigma}_{\boldsymbol\lambda_{jm}})_{1\leq m\leq q}$, and $\text{diag}(\hat{\boldsymbol\Sigma}_{\boldsymbol\gamma_m})=\text{diag}\{\text{diag}(\hat{\boldsymbol\Sigma}_{\boldsymbol\gamma_{mn}^\eta})_{1\leq n\leq q},\text{diag}(\hat{\boldsymbol\Sigma}_{\boldsymbol\gamma_{ml}^x})_{1\leq l\leq Q}\}$, the pointwise covariances for the factor loading parameters are estimated as $$\text{cov}\big(\lambda_{jm}(s),\lambda_{jm}(t)\big)=\mathbf{e}^\top(s)\hat{\boldsymbol{\Sigma}}_{\boldsymbol{\lambda}_{jm}}\mathbf{e}(t),$$ and $$\text{cov}\big(\lambda_{jm}(s,t_1),\lambda_{jm}(s,t_2)\big)=\big(\mathbf{e}^\top(t_1)\otimes\mathbf{e}^\top(s)\big)\hat{\boldsymbol{\Sigma}}_{\boldsymbol{\lambda}_{jm}}\big(\mathbf{e}(t_2)\otimes\mathbf{e}(s)\big),$$   for the concurrent and historical effects of $y_j$ on $\eta_m$, respectively. Similarly, the pointwise covariances for the regression coefficients are estimated as $\text{cov}\big(\gamma_{mn}^\eta(s),\gamma_{mn}^\eta(t)\big)=\mathbf{e}^\top(s)\hat{\boldsymbol{\Sigma}}_{\boldsymbol{\gamma}_{mn}^\eta}\mathbf{e}(t)$ when the effect of $\eta_m$ on $\eta_{n}$ is concurrent, and as $\text{cov}\big(\gamma_{mn}^\eta(s,t_1),\gamma_{mn}^\eta(s,t_2)\big)=\big(\mathbf{e}^\top(t_1)\otimes\mathbf{e}^\top(s)\big)\hat{\boldsymbol{\Sigma}}_{\boldsymbol{\gamma}_{mn}^\eta}\big(\mathbf{e}(t_2)\otimes\mathbf{e}(s)\big)$   when it is historical. Furthermore, in the case when the effect of $\eta_m$ on $x_l$ is smooth, we have $$\text{cov}\big(\gamma_{ml}^x(x_l,t),\gamma_{ml}^x(x_l,t)\big)=\big(\mathbf{e}^\top(t)\otimes\mathbf{h}_l^\top(x_l)\big)\hat{\boldsymbol{\Sigma}}_{\boldsymbol{\gamma}_{ml}^x}\big(\mathbf{e}(t)\otimes\mathbf{h}_l(x_l)\big),$$ and for the constant effect of $\eta_m$ on $x_l$, we can write $$\text{cov}\big(\gamma_{ml}^x(x_{il}),\gamma_{ml}^x(x_{kl})\big)=\mathbf{h}_l^\top(x_{il})\hat{\boldsymbol{\Sigma}}_{\boldsymbol{\gamma}_{ml}^x}\mathbf{h}_l(x_{kl}).$$    Using the pointwise estimate of each parameter accompanied by the estimate of the corresponding pointwise covariance matrix and incorporating them into the hyper-ellipsoid confidence regions provided by the package \texttt{fregion}  \citep{choi2018geometric}, we can derive a confidence band for each functional parameter estimate. The coverage rate (CR) of these proposed confidence bands has been investigated via a simulation study, whose results are detailed in Section \ref{sec:simulation}.

\section{Goodness of fit indices}\label{gofi}

To assess the fit of the FSEM introduced in this paper, we provide two extensions for several absolute and relative goodness of fit indices used for the conventional SEM. The first extension calculates the point-wise fit indices at each time point, and is used to examine the model fit indices at all time points in the domain. The second extension obtains instead the average of the fit indices over the entire time domain. 

Suppose that the degree of freedom is denoted by $df=p(p+1)/2-k$, where $p$ denotes the number of observed variables and  $k$ is the number of estimated parameters. The absolute fit indices at time $t$ include the normed chi-square ($\chi^2/df$) defined as 
\begin{equation*}
    \chi^2(t)=(N-1)F_{ML}(t),
\end{equation*}
with $F_{ML}(t)=\log\vert\boldsymbol\Sigma_t(\boldsymbol{\theta})\vert+\text{tr}(\mathbf{S}_t\boldsymbol\Sigma_t(\boldsymbol{\theta})^{-1})-\log\vert \mathbf{S}_t\vert-p$ denoting the minimum value of the fit function, where $\boldsymbol\Sigma_t(\boldsymbol{\theta})$ is the model implied covariance matrix, and $\mathbf{S}_t$ is the sample  covariance matrix (see Appendix \ref{ap_d}). One can then extend the Root Mean Square Error of Approximation (RMSEA) as
\begin{equation*}
    \text{RMSEA}(t)=\sqrt{\frac{\chi^2(t)-df}{df(N-1)}},
\end{equation*}
and the Standardized Root Mean Square Residual (SRMR) as
\begin{equation*}
 \text{SRMR}(t)=\sqrt{\frac{\sum_{j=1}^p\sum_{k=1}^p(s_{jk}(t)-\hat{\sigma}^2_{jk}(t))^2}{p(p+1)/2}}, 
\end{equation*}
where $s_{jk}(t)$ is the observed covariance of $z_j(t)$ and $z_k(t)$, and $\hat{\sigma}^2_{jk}(t)=\text{cov}(z_j(t),z_k(t))$ denotes the predicted covariance (see Appendix \ref{ap_d}). Finally, the Goodness-of-Fit Index (GFI) can be extended as
\begin{equation*}
 \text{GFI}(t)=1-\frac{\sum_{j=1}^p\sum_{k=1}^p(s_{jk}(t)-\hat{\sigma}^2_{jk}(t))^2}{\sum_{j=1}^p\sum_{k=1}^ps_{jk}(t)^2}.    
\end{equation*}

The relative fit indices compare a $\chi^2$ for the target model to a null (independent) model. The null model is a model in which there are no latent variables.
The relative fit indices used in this paper include an extension to the Comparative Fit Index (CFI)
\begin{equation*}
  \text{CFI}(t)=1-\frac{\chi^2_{\text{model}}(t)-df_{\text{model}}}{\chi^2_{\text{null}}(t)-df_{\text{null}}},  
\end{equation*}
to the Incremental Fit Index (IFI)
\begin{equation*}
\text{IFI}(t)=\frac{\chi^2_{\text{null}}(t)-\chi^2_{\text{model}}(t)}{\chi^2_{\text{null}}(t)-df_{\text{null}}},    
\end{equation*}
and to the Tucker-Lewis Index (TLI)
\begin{equation*}
\text{TLI}(t)=\frac{\chi^2_{\text{null}}(t)/df_{null}-\chi^2_{\text{model}}(t)/df_{model}}{\chi^2_{\text{null}}(t)/df_{(null)}-1}.
\end{equation*}

\section{Simulation studies}\label{sec:simulation}

In this section, we evaluate the performance of the FSEM by carrying out two simulation studies for two separate models. In the first simulation study, described in Section \ref{subsec:simulation1}, we use a simpler model setting, but we experiment across regular and irregular sampling designs, to determine the effect on estimation accuracy and coverage of the number and location of time points per curve.  In the second simulation study, described in Section \ref{subsec:simulation2}, we instead test the performance of the FSEM in a more complex model setup.

\subsection{Simulation 1}\label{subsec:simulation1}

\begin{figure}[tb!]
\centering
				\includegraphics[width=0.8\linewidth]{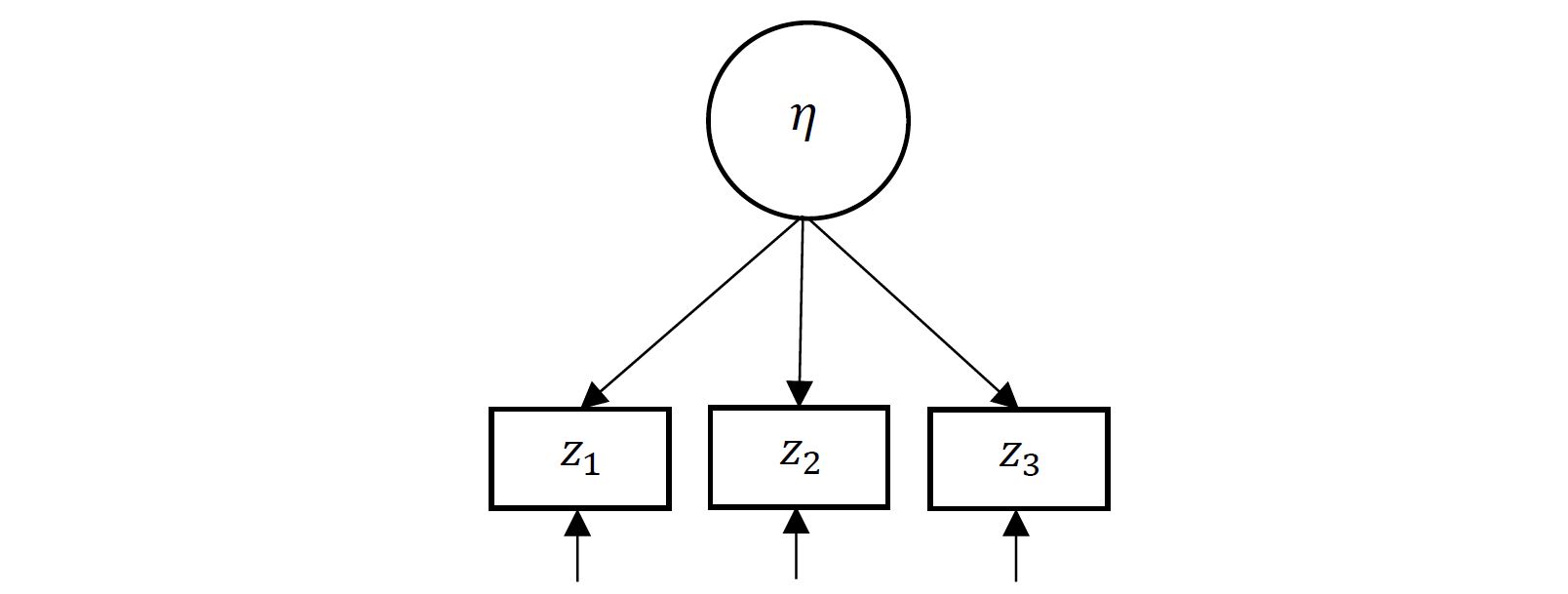} 
	\caption{Path diagram for the FSEM used in Simulation 1, where $p=3$ and $q=1$.}
	\label{path-model2}
\end{figure}

For the first simulation, we consider a latent factor $\eta$ that is identified through $p=3$ indicators in a simple measurement model as illustrated by the path diagram in Figure \ref{path-model2}.  
Suppose the indicators are denoted by $z_1,z_2,z_3$.  The measurement model takes the following form
\begin{align}\label{fac1}
    z_{ij}(t)&=\beta_j(t)+\lambda_j(t)\eta_i(t)+\varepsilon_{ij}(t)+\epsilon_{ijt},\qquad j=1,2,\\\label{fac2}
    z_{i3}(t)&=\beta_3(t)+\int_\tau\lambda_3(s,t)\eta_i(s)ds+\varepsilon_{i3}(t)+\epsilon_{i3t},
\end{align}
where $i=1,2,\ldots,N$.
We use a Karhunen-Lo\`eve expansion to write the residual functions as $\varepsilon_{ij}(t)=\sum_{r\geq 1}\sqrt{\nu_{jr}}u_{ijr}\phi_{jr}(t)$, where $u_{ijr}$ are i.i.d. standard Gaussian random variables and $\nu_{jr}$ and $\phi_{jr}$ are the eigen-values and eigen-functions of the covariance operator of $\varepsilon_j$, respectively. Furthermore, the measurement error terms $\epsilon_{ijt}$ have zero mean Gaussian distributions with variance $\sigma^2_j$. 
Finally,  $\eta_{i}(t)=\sum_{r\geq 1}\sqrt{\mu_{r}}v_{ir}^{(l)}\psi_{r}(t)$, where $\mu_{r}$ and $\psi_{r}$ are the eigen-values and eigen-functions of the covariance operator of $\eta_{i}$, respectively, and $v_{imr}$, are independent standard Gaussian variables. 

We consider two sampling designs in this simulation study. In the first design, the indicators  $z_1,z_2,z_{3}$ are observed at regular time points $t_1,t_2,\ldots,t_{M}$ in the domain.  In the second design, the indicators $z_{ij}$ for $i=1,2,\ldots,N$ and $j=1,2,3$ are observed at the irregular time points $t_{ij1},t_{ij2},\ldots,t_{ijM_{ij}},$ with $M_{ij}\sim U(\{a(1/2)^{M}, a(1/2)^{M-1}, \ldots, a(1/2)^1\})$ where $a=2^M/(2^M-1)$ is the normalizing constant, and with $t_{ijk}\sim U([0,1])$ i.i.d. for $k=1,2,\ldots,M_{ij}$ (in increasing order in $k$); $U(A)$ denotes the uniform distribution over the (continuous or discrete) set $A$.
We set the functional intercepts and regression coefficients in the measurement model respectively as $\beta_j(t)=jt^2$,   $\lambda_j(t)=1+\frac{1}{2}\sin(\pi \sqrt{j}t/2)$ for $j=1,2$, and $\lambda_3(t)=1+\frac{1}{2}\cos(\pi \sqrt{3}(s+t))$.
The covariance structures for residual functions are constructed as $\nu_{jr}=k\rho^{(r-1)},$ where $k>0$ is a fixed value and ${0< \rho<1}$, and for $r\geq 0$, we set the eigen-functions as $\phi_{j(2r)}(t)=\sqrt{2}\sin(2\pi trj)$ and $\phi_{j(2r+1)}(t)=\sqrt{2}\cos(\pi t(2r+1)j)$.
Moreover, we suppose that the covariance function for the latent variable $\eta$ is defined as $K^\eta(s,t)=\exp\{-(t-s)^2\},$ which belongs to the family of Mat\'en covariance functions. 
 The measurement error variances are chosen using the Signal-to-Noise Ratio (SNR), defined by
 \begin{equation*}
  SNR=\frac{\int_0^1 K^{\varepsilon}_j(t,t)dt}{\sigma^2_j},   
 \end{equation*}
where $K^{\varepsilon}_j$ denotes the kernel function for the covariance operator of $\varepsilon_{j}$.

We assume that only the first $8$ eigen-values are non-zero in the Karhunen-Lo\`eve expansions of the residual functions and of the latent variable. The decay rate of eigen-values is identified by choosing the parameter $\rho$. Values of $\rho$ approaching zero (one) correspond to smoother (less smooth) residual functions. Consequently, by changing the value of $\rho$, we can change the complexity of the covariance structure.

We run the simulations for $k=1$, $\rho=0.3$, $SNR=4,$ and all the following combinations of sampling designs and model parameters:
\begin{itemize}
	\item $M=10, 20$ for a small and large number of time points, respectively.
	\item $N=50,  100$ for a small and large sample size, respectively.
    \item Regular and irregular sampling designs. 
\end{itemize}
Moreover, we set $J=10, 12$ for $M=10, 20$, respectively.

The accuracy of the estimates is evaluated by calculating the Mean Squared Error (MSE) of the estimators by using an MCMC algorithm with 200 iterations. The results are reported in Table \ref{t.fac1}. The Coverage Rate (CR) of the 95\% confidence bands computed as described in Section \ref{conBan} is reported in Table \ref{t.cov1}. 

As shown in Table \ref{t.fac1}, the MSE of the estimates is generally quite low, indicating a good accuracy of the FSEM; the MSE decreases with an increase in $N$, and it sometimes increases in the IR setting, as expected.
Additionally, the CRs in Table \ref{t.cov1} indicate a good coverage both for the regular and irregular designs, with a slightly better performance in the former case (and for larger $N$), as expected. All in all, from these tables we can conclude that the FSEM provides a good overall fit, yielding accurate parameter estimates and well-calibrated CRs.

\begin{table}[t]
\centering
	\caption{Results of Simulation 1. MSE of parameter estimates for the measurement models defined in (\ref{fac1})  and  (\ref{fac2}), in the case of regular (R) and irregular (IR) designs.}
	\label{t.fac1}
 {\footnotesize
\begin{tabular}{ccccccccccc}
\hline
 \multirow{9}{*}{FM(1)} & \multirow{5}{*}{R}  & \textbf{N}           & \textbf{M} & $\boldsymbol{\beta}_1$ & $\boldsymbol{\lambda}_1$ & $\boldsymbol{\phi}_{11}$ &  $\boldsymbol{\nu}_{11}$ &  $\boldsymbol{\sigma}_1^2$ \\\hline
                        &                     & \multirow{2}{*}{50}  & 10         & 0.053                  & 0.024                    & 0.022                    & 0.110                   & \phantom{<<<}0.004                     \\
                       &                     &                      & 20         & 0.051                  & 0.028                    & 0.021                    & 0.092                   & \textless{}0.001                     \\
                       &                     & \multirow{2}{*}{100} & 10         & 0.027                  & 0.012                    & 0.018                    & 0.064                   & \phantom{<<<}0.003                     \\
                       &                     &                      & 20         & 0.031                  & 0.021                    & 0.018                    & 0.070                   & \textless{}0.001                     \\
                       & \multirow{4}{*}{IR} & \multirow{2}{*}{50}  & 10         & 0.048                  & 0.037                    & 0.039                    & 0.053                   & \textless{}0.001                     \\
                       &                     &                      & 20         & 0.045                  & 0.021                    & 0.036                    & 0.052                   & \textless{}0.001                     \\
                       &                     & \multirow{2}{*}{100} & 10         & 0.024                  & 0.021                    & 0.026                    & 0.046                   & \textless{}0.001                     \\
                       &                     &                      & 20         & 0.024                  & 0.010                    & 0.024                    & 0.054                   & \textless{}0.001                     \\
      \\ \hline

\multirow{9}{*}{FM(2)} & \multirow{5}{*}{R}  & \textbf{N}           & \textbf{M} & $\boldsymbol{\beta_2}$ & $\boldsymbol{\lambda}_2$ & $\boldsymbol{\phi}_{21}$ &  $\boldsymbol{\nu}_{21}$ &  $\boldsymbol{\sigma}_2^2$ \\\hline
                         &                     & \multirow{2}{*}{50}  & 10         & 0.048                  & 0.025                    & 0.026                    & 0.043                   & \phantom{<<<}0.003                     \\
                       &                     &                      & 20         & 0.048                  & 0.026                    & 0.021                    & 0.038                   & \textless{}0.001                     \\
                       &                     & \multirow{2}{*}{100} & 10         & 0.025                  & 0.013                    & 0.012                    & 0.020                   & \phantom{<<<}0.002                     \\
                       &                     &                      & 20         & 0.028                  & 0.016                    & 0.009                    & 0.023                   & \textless{}0.001                     \\
                       & \multirow{4}{*}{IR} & \multirow{2}{*}{50}  & 10         & 0.042                  & 0.031                    & 0.044                    & 0.030                   & \textless{}0.001                     \\
                       &                     &                      & 20         & 0.034                  & 0.020                    & 0.032                    & 0.023                   & \textless{}0.001                     \\
                       &                     & \multirow{2}{*}{100} & 10         & 0.021                  & 0.016                    & 0.028                    & 0.013                   & \textless{}0.001                     \\
                       &                     &                      & 20         & 0.018                  & 0.010                    & 0.027                    & 0.013                   & \textless{}0.001                     \\
        \\ \hline

\multirow{9}{*}{FM(3)} & \multirow{5}{*}{R}  & \textbf{N}           & \textbf{M} & $\boldsymbol{\beta}_3$ & $\boldsymbol{\lambda}_3$ & $\boldsymbol{\phi}_{31}$ &  $\boldsymbol{\nu}_{31}$ & $\sigma_3^2$             \\\hline
                       &                     & \multirow{2}{*}{50}  & 10         & 0.024                  & 0.081                    & 0.259                    & 0.191                   & \phantom{<<<}0.003                     \\
                       &                     &                      & 20         & 0.024                  & 0.061                    & 0.025                    & 0.052                   & \phantom{<<<}0.001                     \\
                       &                     & \multirow{2}{*}{100} & 10         & 0.015                  & 0.065                    & 0.232                    & 0.154                   & \phantom{<<<}0.002                     \\
                       &                     &                      & 20         & 0.016                  & 0.048                    & 0.019                    & 0.023                   & \phantom{<<<}0.001                     \\
                       & \multirow{4}{*}{IR} & \multirow{2}{*}{50}  & 10         & 0.021                  & 0.046                    & 0.039                    & 0.024                   & \phantom{<<<}0.003                     \\
                       &                     &                      & 20         & 0.018                  & 0.042                    & 0.021                    & 0.023                   & \textless{}0.001                     \\
                       &                     & \multirow{2}{*}{100} & 10         & 0.012                  & 0.037                    & 0.019                    & 0.014                   & \phantom{<<<}0.004                     \\
                       &                     &                      & 20         & 0.010                  & 0.030                    & 0.011                    & 0.011                   & \phantom{<<<}0.001                    
        \\ 
\hline     
\end{tabular}
}
\end{table}

\begin{table}[t]
\centering
	\caption{Results of Simulation 1. CR of the confidence bands for the regression coefficient estimates in the case of regular (R) and irregular (IR) designs.}
	\label{t.cov1}
 {\footnotesize
\begin{tabular}{cccccc}
\hline
\textbf{design}     & \textbf{N}           & \textbf{M} & $\boldsymbol\lambda_1$ & $\boldsymbol\lambda_2$ & $\boldsymbol\lambda_3$ \vspace{0.2em} \\\hline
\multirow{4}{*}{R}  & \multirow{2}{*}{50}  & 10         & 0.951                  & 0.957                  & 0.962                  \\
                    &                      & 20         & 0.940                  & 0.954                  & 0.935                  \\
                    & \multirow{2}{*}{100} & 10         & 0.960                  & 0.970                  & 0.914                  \\
                    &                      & 20         & 0.906                  & 0.956                  & 0.919                  \\
\multirow{4}{*}{IR} & \multirow{2}{*}{50}  & 10         &    0.912                    &      0.923                  &  0.934                      \\
                    &                      & 20         &   0.932                     &    0.935                    & 0.937                       \\
                    & \multirow{2}{*}{100} & 10         &  0.914                       &    0.928                    &   0.914                     \\
                    &                      & 20         &   0.944                     &    0.950                    & 0.916\\\hline                      
\end{tabular}
}
\end{table}

\subsection{Simulation 2}\label{subsec:simulation2}

\begin{figure}[tb!]
\centering
				\includegraphics[width=0.8\linewidth]{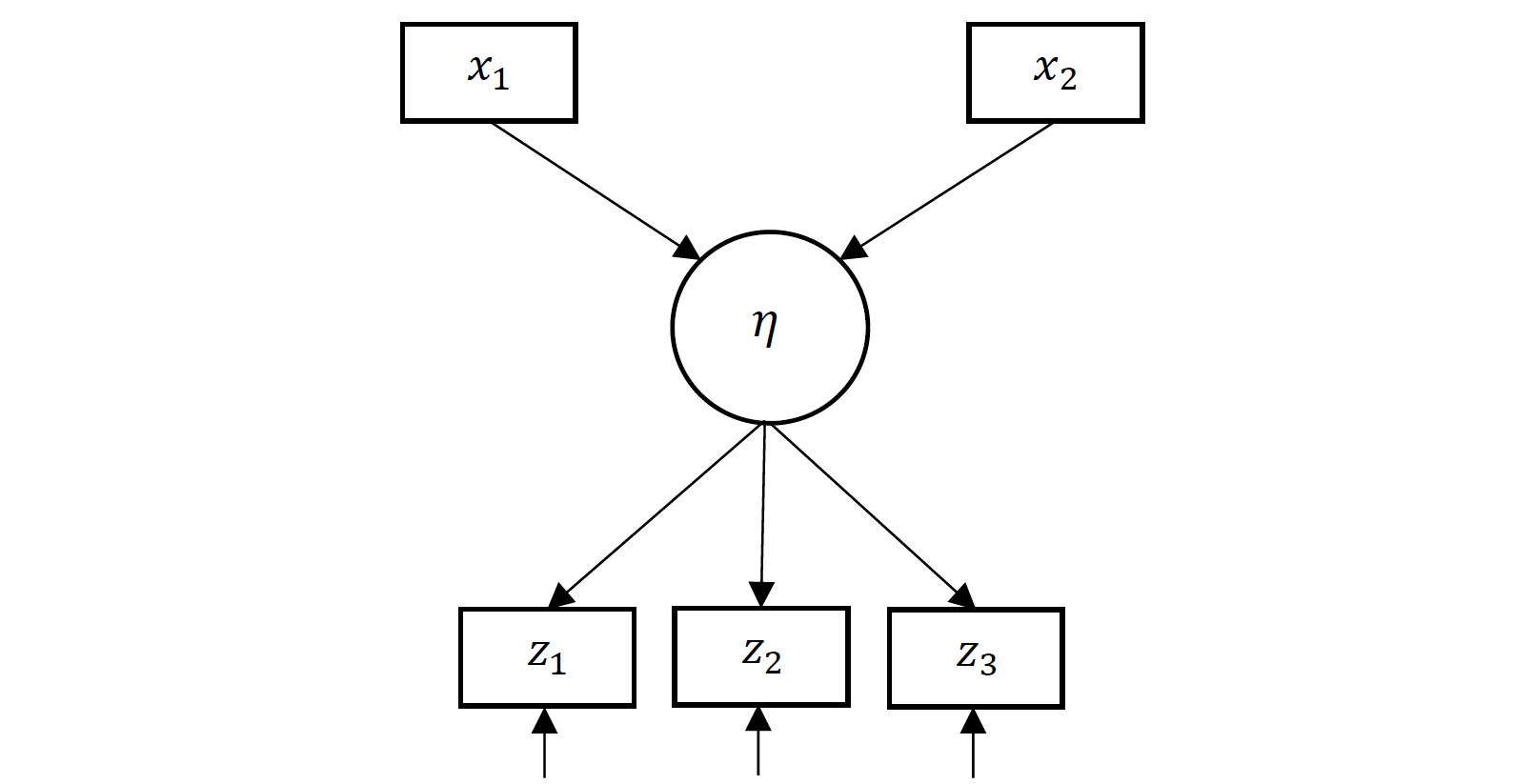} 
	\caption{Path diagram for the FSEM used in Simulation 2, where $p=3$ and $q=1$.}
	\label{path-model1}
\end{figure}

For the second simulation study, we implement a scenario that resembles the model that will be fitted to the real data example detailed in Section \ref{sec:casestudy}. 
For this purpose, consider an FSEM with $p=3$ indicators and $q=1$ latent factor, together with $Q=2$ observed scalar-valued explanatory variables (see the path diagram illustrated in Figure \ref{path-model1}). 
Suppose the latent variable $\eta$ is measured through indicators $z_1,z_2,z_{3}$ via the measurement model defined as
\begin{align}\label{fac3}
    z_{ij}(t)&=\beta_j(t)+\lambda_j(t)\eta_{i}(t)+\varepsilon_{ij}(t)+\epsilon_{ijt},\qquad j=1,2,3,
\end{align}
for $i=1,2,\ldots,N$.
Further, for the structural part of the FSEM, we have 
\begin{align}\label{sem1}
    \eta_{i}(t)&=\gamma_{1}^x(t)x_{i1}(t)+\gamma_{2}^x(t)x_{i2}(t)+\zeta_{i}(t).
\end{align}

We assume that the indicators $z_{ij}$ for $i=1,2,\ldots,N$ and $j=1,2,3$ are observed at regular missing-at-random time points $t_{ijk}=k/M, k=1,2,\ldots,M$. The indicator $z_{ij}$ may be missing at $t_{ijk}$ according to a Bernoulli distribution with success probability $p=0.12$.  For generating data from the FSEM represented in Figure \ref{path-model1}, the functional coefficients in the measurement models are chosen as $\beta_j(t)=jt$, $\lambda_j(t)=1+\frac{1}{2}\cos(\pi \sqrt{j}t)$ for $j= 1,2,3$, and
for the structural models, we choose the parameters $\gamma_{1}^x(t)=t^2$ and $\gamma_{2}^x(t)=2t^2$. 

 The covariance structures for the residual functions are constructed using 
 the Karhunen-Lo\`eve expansions with the same eigen-values and eigen-functions used in Simulation 1. The unique factors are constructed 
using covariance functions belonging to the Matérn family. We use the Matérn covariance functions $K_{m}^{\zeta}(s,t)=\exp{(-2m\vert t-s\vert)}$, for $m=1,2,3$.

The values for $\sigma^2_j$, $k$, and  $\rho$  are set as in Simulation 1. The remaining parameters are selected as follows:  $J=6$, $M=8$, and $N=100$. To assess the accuracy of the estimates, the MSEs of the estimators for the functional parameters, including factor loadings and regression coefficients, and the CRs of 95\% confidence bands, are reported in Table \ref{t.sim2}. The table shows a pretty good accuracy and coverage of the model for nearly all parameters,  with the exception of $\boldsymbol{\gamma}_2^x$, which shows slightly worse MSE and CR.

\begin{table}[t]
\centering
	\caption{Results of Simulation 2. MSE and CR of parameter estimates for the measurement and the structural models defined in (\ref{fac3})  and  (\ref{sem1}), respectively.}
	\label{t.sim2}
 {\footnotesize
\begin{tabular}{cccccc}
\vspace{0.05cm}
    & $\boldsymbol{\lambda}_1$ & $\boldsymbol{\lambda}_2$ & $\boldsymbol{\lambda}_3$ & $\boldsymbol{\gamma}_1^x$ & $\boldsymbol{\gamma}_2^x$ \\\hline
MSE & 0.036                    & 0.021                    & 0.037                    & 0.026                     & 0.078                     \\
CR  &  0.943                        &     0.953                     &    0.947                      &    0.951                       &        0.918                  
\end{tabular}
}
\end{table}

\section{Application of the FSEM to the analysis of the Health and Retirement Study (HRS) data}\label{sec:casestudy}
In this section, a subset of the dataset extracted from the  HRS  \citep{HRS2024} is analyzed to demonstrate the performance of the FSEM. 
This analysis uses Early Release data from the Health and Retirement Study, (Public Survey Data), sponsored by the National Institute on Aging (grant number NIA U01AG009740) and conducted by the University of Michigan. These data have not been cleaned and may contain errors that will be corrected in the Final Public Release version of the dataset.

For the purposes of this analysis, a random sample of 400 participants who were involved in eight waves of data collection, from 2008 to 2022, was selected. The main goal of the analysis is to investigate the General Factor of Personality (GFP) as a latent factor identified by the Big Five personality traits. 

Covariates available in the HRS include `sex' (0=female, 1=male), `cancer' (0=no cancer, 1=having been diagnosed with any type of cancer), `diabetes' (0=no diabetes, 1=having been diagnosed with diabetes), `heart conditions' (0=no heart disease, 1=having been diagnosed with any type of heart disease), and the Big Five personality traits including `neuroticism', `extraversion', `openness', `agreeableness', and `conscientiousness'. 
The Big Five personality traits are regularly measured at eight measurement time points for each participant, but the data may be missing for some participants in some years.

The main goal of the analysis is to address the following research questions: (1) how does the structure of GFP, as a latent variable that explains the functional covariance between the Big Five personality traits, evolve in time (i.e., how the factor loadings estimated for GFP change across time)? (2) how do the personality factors constitute GFP? (3) how does GFP score differ between men and women over a time window from 2008 to 2022? And finally (4) how does GFP change if a person has been diagnosed with any type of cancer (with respect to no diagnosis)?

A matching procedure was employed to select a balanced sample of 400 participants, divided equally into two groups: those diagnosed with any type of cancer and those without any diagnosis. Initially, 200 participants with a cancer diagnosis were randomly selected. For each participant in the cancer group, a corresponding participant without a cancer diagnosis was chosen. The matching process ensures that each pair is comparable in terms of sex, diabetes status, and cardiovascular conditions. This approach aimed to control for these confounding variables, thereby enhancing the validity of the comparative analysis between the two groups.

 The latent function GFP is identified through the Big Five personality traits. Here, each of the five traits has been considered as an observed variable by computing the arithmetic mean of its items (per measurement time points), and it thus can be considered as an `item parcel' as in \cite[p. 168]{newsom2015longitudinal}. Although item parcels have various potential drawbacks, they are used here for pedagogical convenience to circumvent a more complex second-order factor model.
Item parcels are generally less problematic when their components can be demonstrated to originate from a common factor that exhibits a strong relationship with the overarching factor
 \citep{bandalos2001item}.

An FSEM with $q=1$ latent variable, $p=5$ indicators, and different covariates is fitted to the data to assess the relationship between the observed covariates `sex' and `cancer' at baseline, and the latent response variable GFP. The functional latent variable GFP for the $i^{\text{th}}$ participant is denoted by $\text{GFP}_i,$ and its corresponding indicators including `neuroticism', `extraversion', `openness', `agreeableness', and `conscientiousness' are represented respectively by $z_{ij}$ for $j=1,2,\ldots,5$ with $i=1,2,\ldots,400$. Moreover, the variables $\text{G}_i$, and $\text{C}_i$ indicate respectively the Sex and Cancer covariates for the $i^{\text{th}}$ participant. 

The following FSEM model is fitted to the dataset to obtain smooth estimates for the factor loadings in the measurement model and for the regression coefficients in the structural model:
\begin{align}
    z_{ij}(t_{k})&=\beta_j(t_k)+\lambda_j(t_k)\text{GFP}_i(t_k)+\varepsilon_{ij}(t_k)+\epsilon_{ijt_k},\\
    \label{model:HRS}
    \text{GFP}_i(t_k)&=\gamma^x_1(t_k)\text{G}_i+\gamma^x_2(t_k)\text{C}_i+\gamma^x_3(t_k)\text{G}_i\text{C}_i+\zeta_i(t_k),
\end{align}
for $k=1,2,\ldots,8$, where $\varepsilon_{ij}$ and $\zeta_i$ stand for the residual function and the unique factor, and $\epsilon_{ijt_k}$ represents the measurement error. The main effects of `diabetes' and `heart condition' and their two-way interactions with other covariates are also included in the structural part \eqref{model:HRS} of the model (not shown for the sake of simplicity). For some participants, missing values occur at some time points $t_k; k=1,2,\ldots,8$. We assume a regular design with equally spaced, Missing Completely-At-Random (MCAR) time points for the study since Little’s test \citep{li2013little} did not reject the hypothesis of MCAR. 

\begin{figure}[tb!]
\begin{minipage}{\dimexpr 0.2\linewidth-3cm\relax}
		\rotatebox{90}{$~$}
		
	\end{minipage}%
	\begin{minipage}{\dimexpr 0.5cm\relax}
		\rotatebox{90}{\scriptsize $\text{tol}_{coef}$}
		
	\end{minipage}%
	\begin{minipage}{\dimexpr\linewidth-0.5cm\relax}%
			\begin{minipage}{0.41\linewidth}
				\centering
				\includegraphics[width=\textwidth]{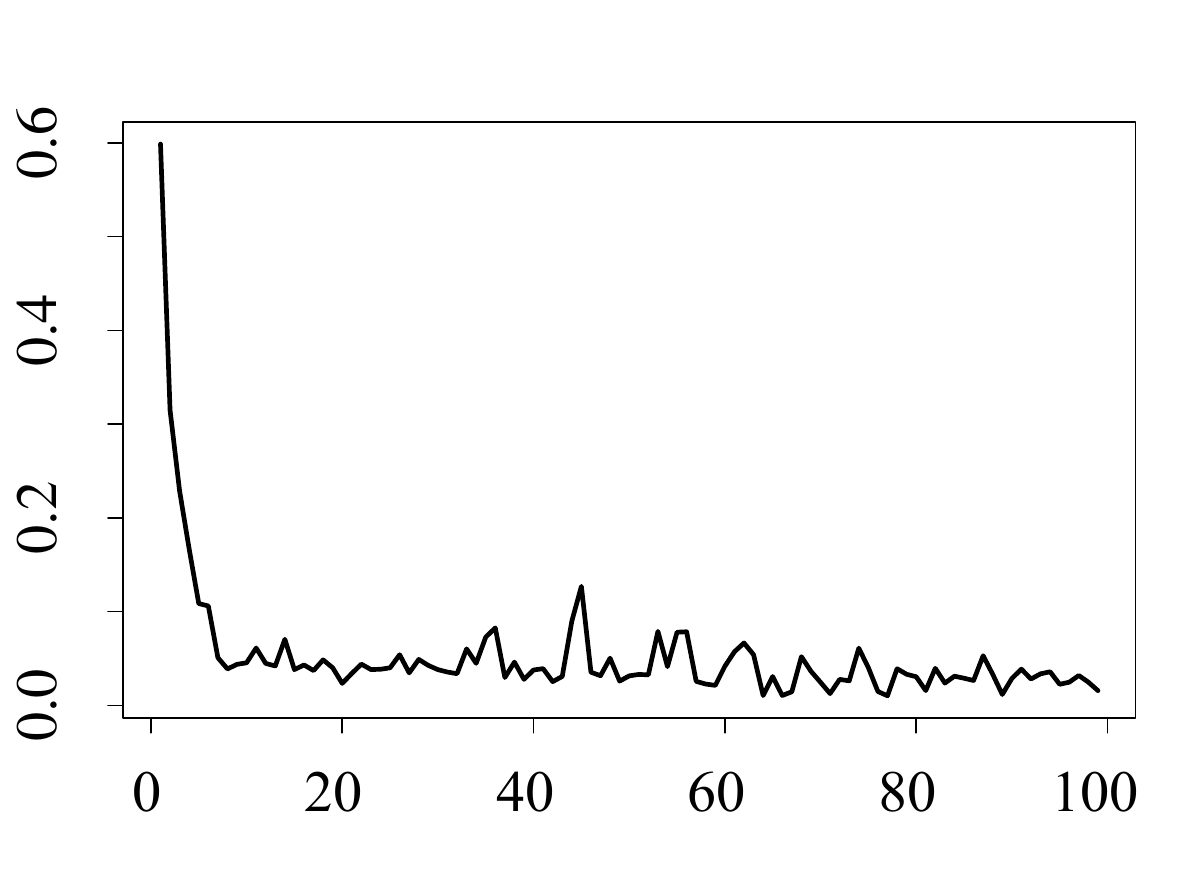}	
			\end{minipage}
   \hspace{-9cm}
			\begin{minipage}{0.63\linewidth}
		\rotatebox{90}{$~$}
		
	\end{minipage}%
			\begin{minipage}{\dimexpr 0.5cm\relax}
		\rotatebox{90}{\scriptsize$\text{tol}_{\sigma^2}$}
	\end{minipage}%
			\begin{minipage}{0.5\linewidth}
				\centering
				\includegraphics[width=\textwidth]{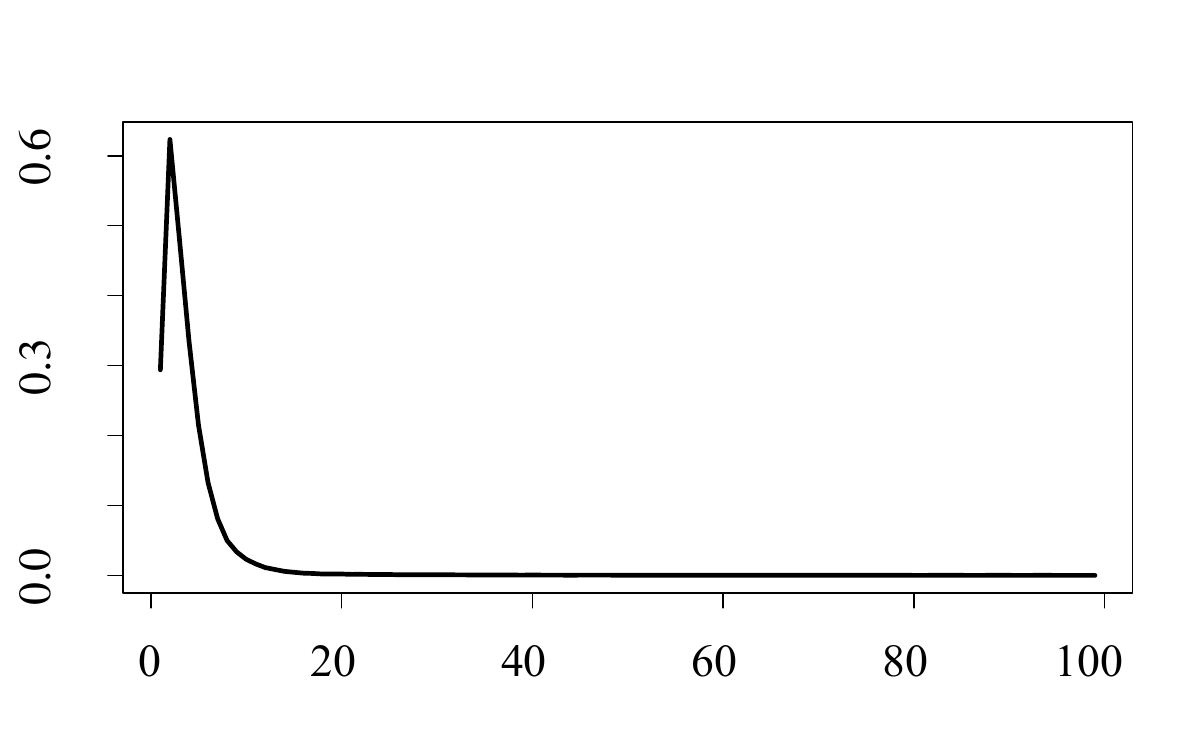} 	
			\end{minipage}
			\begin{minipage}{\dimexpr 0.2\linewidth-1.25cm\relax}
		\rotatebox{90}{$~$}
		
	\end{minipage}%
	\end{minipage}%
	
		\begin{minipage}{1\linewidth}
				\centering
				\hspace{-2em}\footnotesize{EM iteration}
			\end{minipage}
	\caption{Results of the HRS data analysis. Convergence of parameters estimation along the EM algorithm iterations. Left panel: total tolerance of the functional parameters, as estimated via the sum of the tolerances of factor loadings and regression coefficients. Right panel: total tolerance of the measurement errors, as estimated via the sum of the tolerances of the measurement error for each of the indicators.}

	\label{convergence}
\end{figure}

The FSEM model defined above is fitted to the data with $J=6$ basis functions. Figure \ref{convergence} demonstrates convergence of the fitting procedure when inspecting the sum of tolerances of the estimated functional parameters including factor loadings and regression coefficients as well as the sum of the tolerances of the estimated measurement error variances $\sigma^2_j; j=1,2,\ldots,5$. 

\begin{figure}[tb!]
\begin{minipage}{\dimexpr 0.2\linewidth-3cm\relax}
		\rotatebox{90}{$~$}
		
	\end{minipage}%
       \hspace{-0.5cm}
	\begin{minipage}{0.5cm}
\rotatebox{90}{\scriptsize$\lambda_1(t)$}

	\end{minipage}%
	\begin{minipage}{\dimexpr\linewidth-0.5cm\relax}%
			\begin{minipage}{0.5\linewidth}
				\centering
				\includegraphics[width=\textwidth]{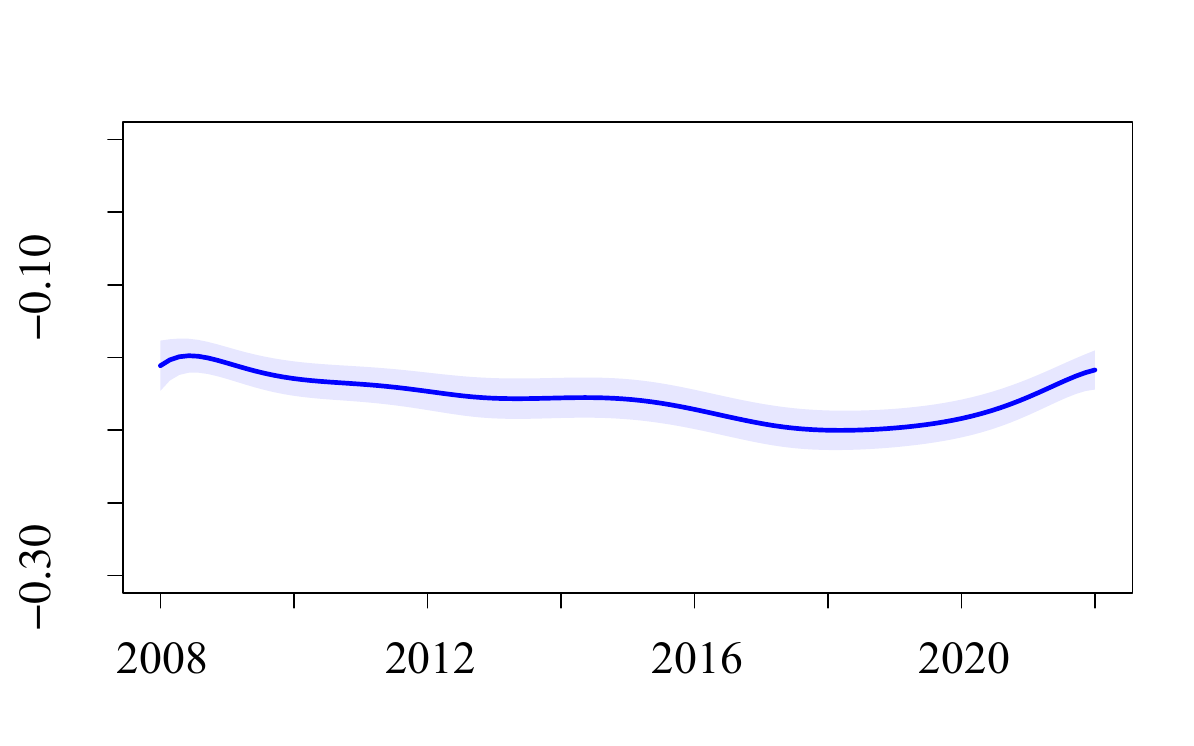}	
			\end{minipage}
      \hspace{-0.5cm}
			\begin{minipage}{0.5cm}
		\rotatebox{90}{$~$}
		
	\end{minipage}%
			\begin{minipage}{\dimexpr 0.5cm\relax}
		\rotatebox{90}{\scriptsize$\lambda_2(t)$}
	\end{minipage}%
			\begin{minipage}{0.5\linewidth}
				\centering
				\includegraphics[width=\textwidth]{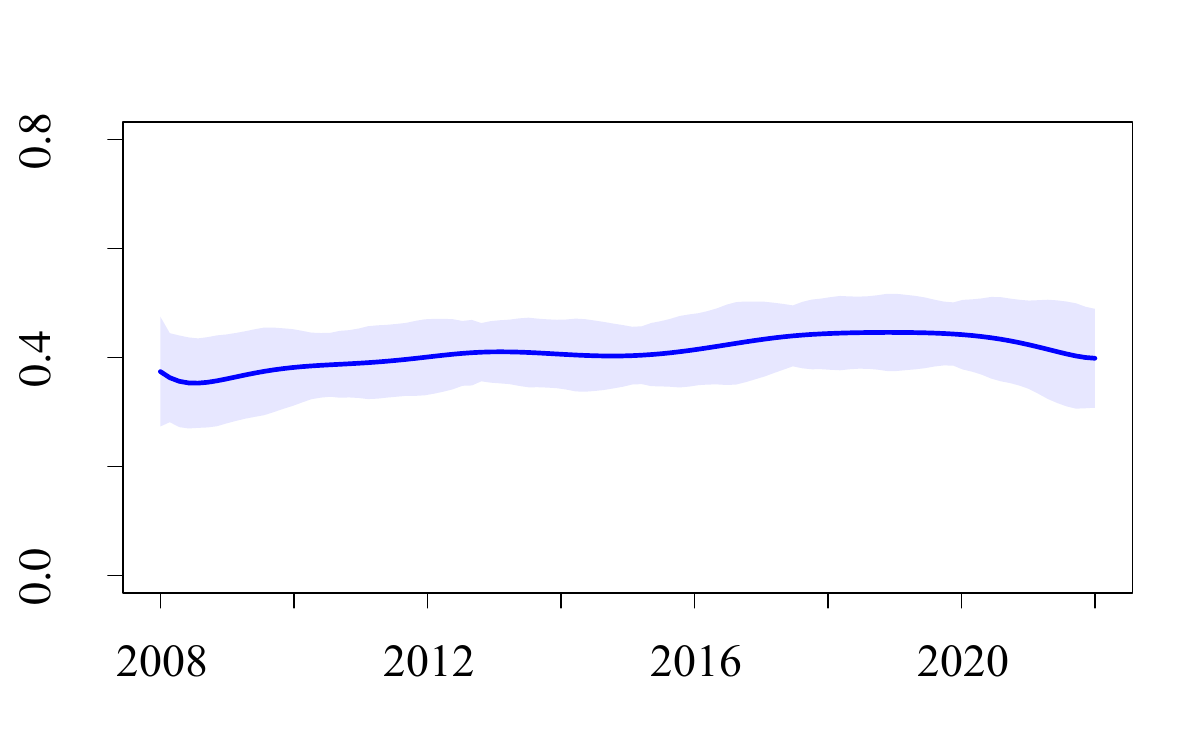} 
			\end{minipage}
			\begin{minipage}{\dimexpr 0.2\linewidth-1.25cm\relax}
		\rotatebox{90}{$~$}
		
	\end{minipage}%
	\end{minipage}%

\begin{minipage}{\dimexpr 0.2\linewidth-3cm\relax}
		\rotatebox{90}{$~$}
		
	\end{minipage}%
        \hspace{-0.5cm}
	\begin{minipage}{\dimexpr 0.5cm\relax}
		\rotatebox{90}{\scriptsize$\lambda_3(t)$}
		
	\end{minipage}%
	\begin{minipage}{\dimexpr\linewidth-0.5cm\relax}%
			\begin{minipage}{0.5\linewidth}
				\centering
				\includegraphics[width=\textwidth]{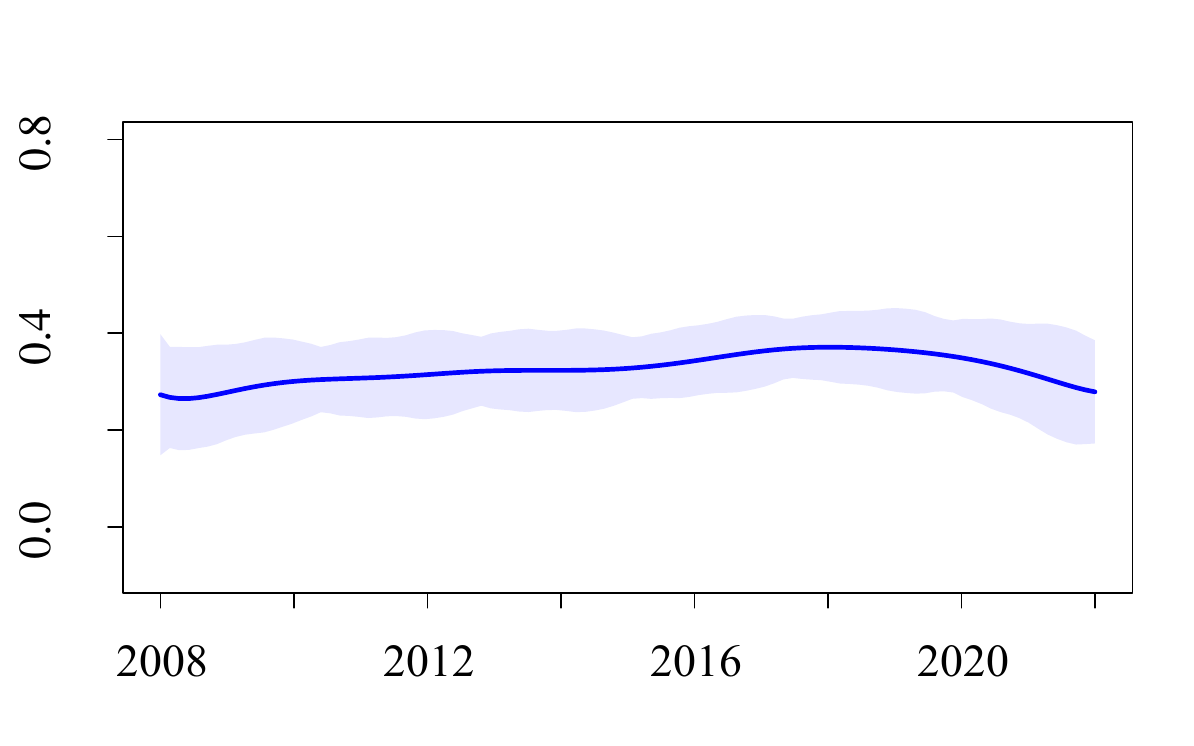}		
			\end{minipage}
            \hspace{-0.5cm}
			\begin{minipage}{0.5cm}
		\rotatebox{90}{$~$}
		
	\end{minipage}%
			\begin{minipage}{\dimexpr 0.5cm\relax}
		\rotatebox{90}{\scriptsize$\lambda_4(t)$}
	\end{minipage}%
			\begin{minipage}{0.5\linewidth}
				\centering
				\includegraphics[width=\textwidth]{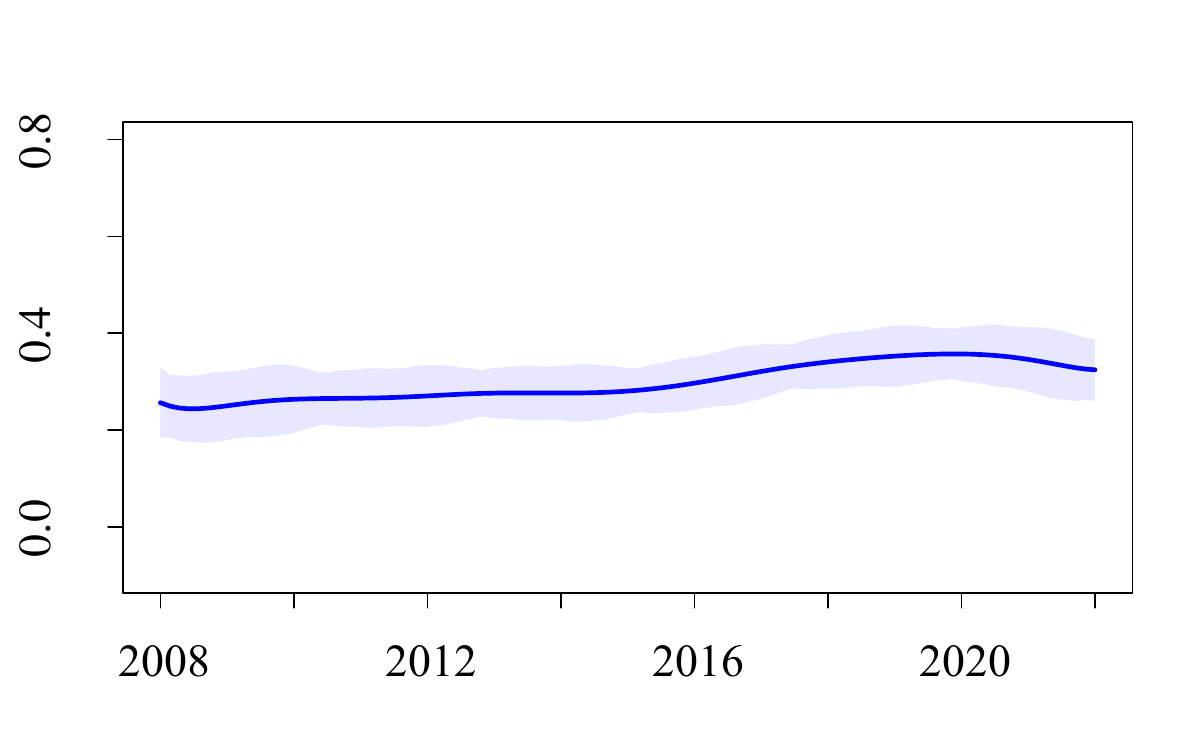} 
			\end{minipage}
			\begin{minipage}{\dimexpr 0.2\linewidth-1.25cm\relax}
		\rotatebox{90}{$~$}
		
	\end{minipage}%
	\end{minipage}%

 \begin{minipage}{\dimexpr 0.2\linewidth-1.25cm\relax}
		\rotatebox{90}{$~$}
		
	\end{minipage}%
        \hspace{-0.5cm}
 \hspace{6em}
	\begin{minipage}{\dimexpr 0.5cm\relax}
		\rotatebox{90}{\scriptsize$\lambda_5(t)$}
		
	\end{minipage}%
	\begin{minipage}{\dimexpr\linewidth-0.5cm\relax}%
			\begin{minipage}{0.5\linewidth}
				\centering
				\includegraphics[width=\textwidth]{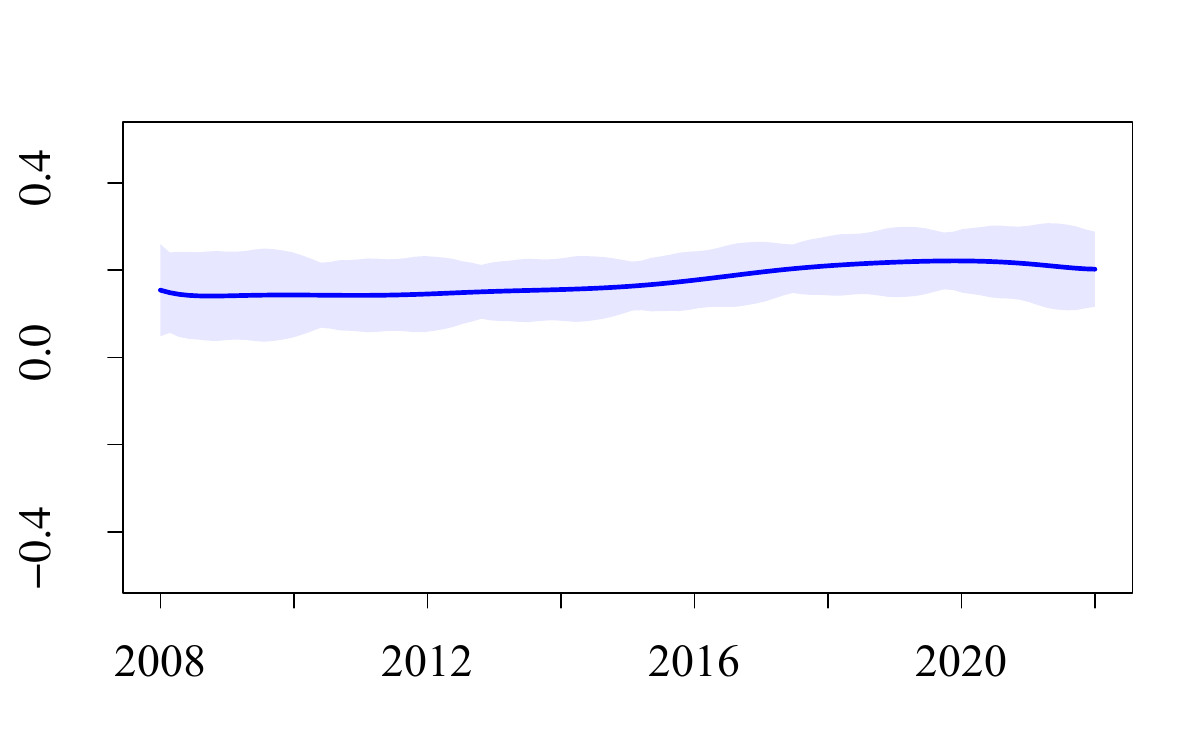}	
			\end{minipage}
			\begin{minipage}{0.5cm}
		\rotatebox{90}{$~$}
		
	\end{minipage}%
			
	\end{minipage}%
 \vspace{1em}
		\begin{minipage}{1\linewidth}
				\centering
				\hspace{0.5cm}\footnotesize{Year}
			\end{minipage}
	\caption{Results of the HRS data analysis. Estimated factor loadings (blue solid lines) with the corresponding 95\% confidence bands (light blue area). Left and right upper panels correspond respectively to the indicators `neuroticism' and `extraversion' and left and right middle panels correspond to `openness' and `agreeableness', respectively. Additionally, lower panel corresponds to `conscientiousness'. }

	\label{factorLoading}
\end{figure}

\begin{figure}[tb!]
\begin{minipage}{\dimexpr 0.2\linewidth-3cm\relax}
		\rotatebox{90}{$~$}
		
	\end{minipage}%
       \hspace{-0.5cm}
	\begin{minipage}{0.5cm}
\rotatebox{90}{\scriptsize$\gamma_1^x(t)$}

	\end{minipage}%
	\begin{minipage}{\dimexpr\linewidth-0.5cm\relax}%
			\begin{minipage}{0.5\linewidth}
				\centering
				\includegraphics[width=\textwidth]{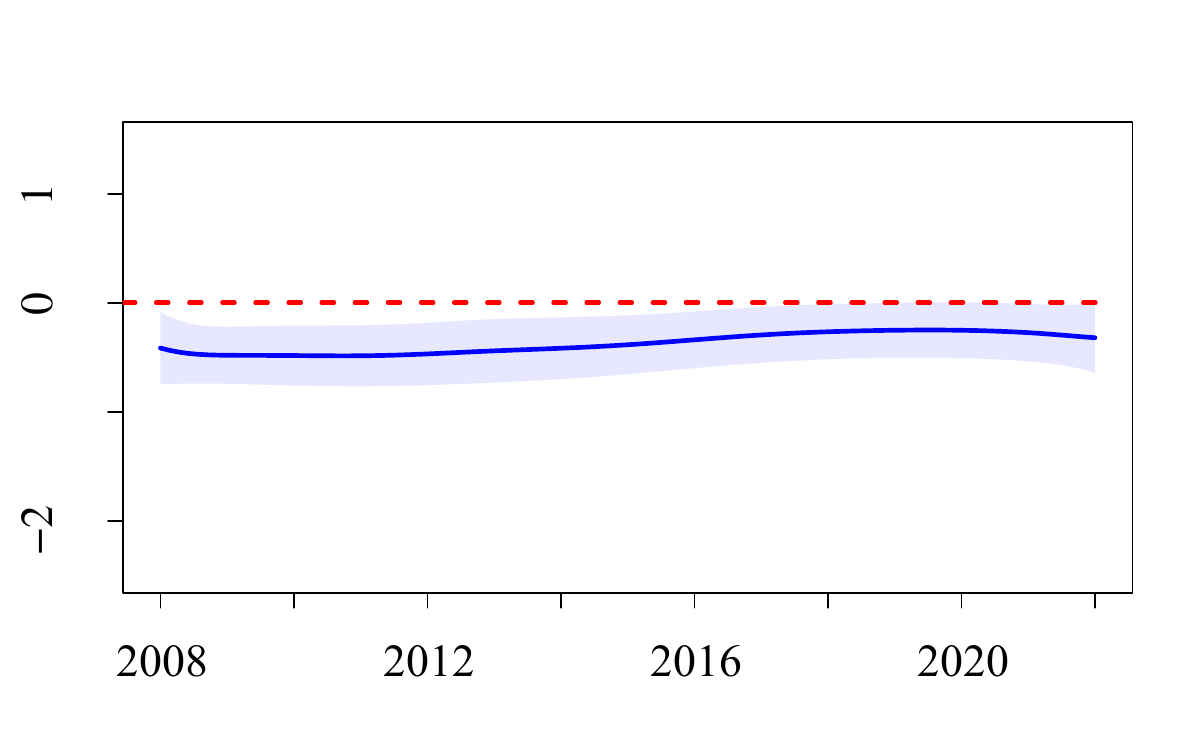}	
			\end{minipage}
      \hspace{-0.5cm}
			\begin{minipage}{0.5cm}
		\rotatebox{90}{$~$}
		
	\end{minipage}%
			\begin{minipage}{\dimexpr 0.5cm\relax}
		\rotatebox{90}{\scriptsize$\gamma_2^x(t)$}
	\end{minipage}%
			\begin{minipage}{0.5\linewidth}
				\centering
				\includegraphics[width=\textwidth]{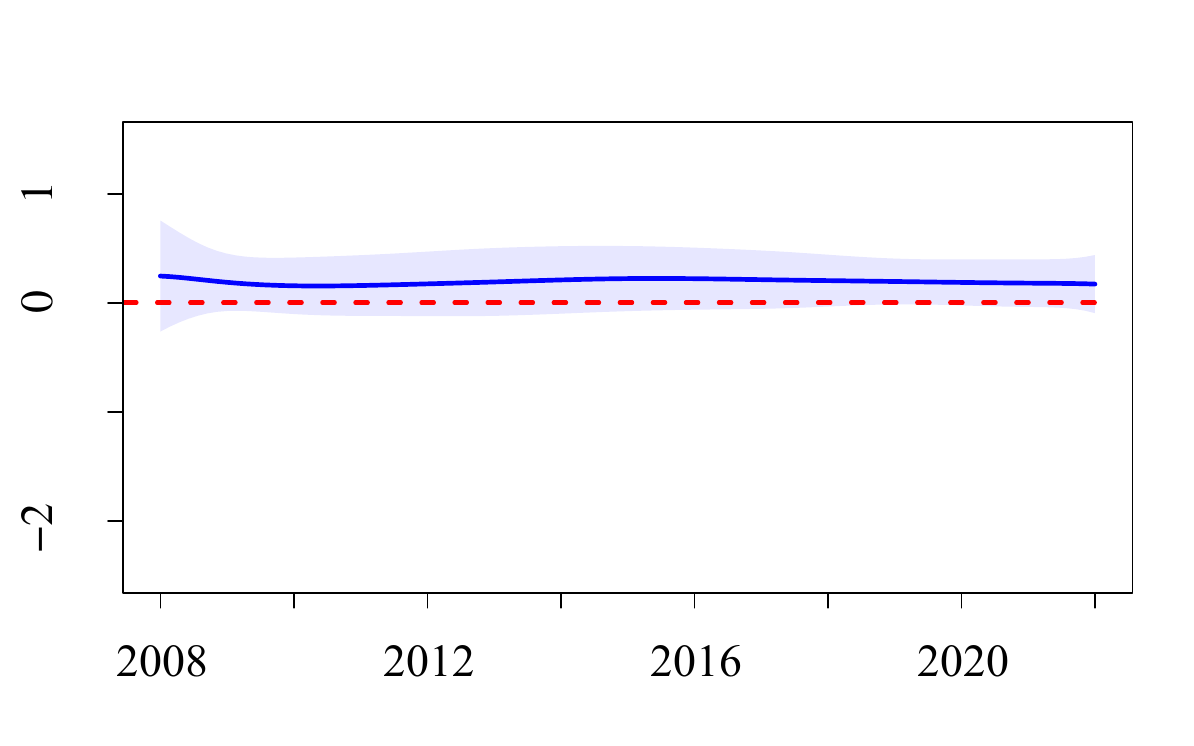} 
			\end{minipage}
			\begin{minipage}{\dimexpr 0.2\linewidth-1.25cm\relax}
		\rotatebox{90}{$~$}
		
	\end{minipage}%
	\end{minipage}%

	\caption{Results of the HRS data analysis. Estimated regression coefficients with the corresponding 95\% confidence bands.  Left panel: main effect of sex;  right panel: main effect of cancer status.}
	\label{regressionCoef}
\end{figure}

The smooth estimates of the factor loadings are shown in Figure \ref{factorLoading}, while the regression coefficients for the main effects of `sex' and `cancer' (their two-way interaction was not significant, therefore not shown) are illustrated in Figure \ref{regressionCoef}. The main effects of the `diabetes' and `heart conditions' and their two-way interactions are not shown, as they were not significant due to their 95\% confidence bands including zero throughout the domain.

According to Figure \ref{factorLoading}, the factor loadings of the indicators  `extraversion', `openness', `agreeableness', and `conscientiousness' are all positive, and they seem to be constant in time, as their confidence bands resemble a constant line. In contrast, the factor loading of `neuroticism' is overall negative, but it changes slightly over time, with its amplitude increasing until the end of 2019 and decreasing afterward. According to the sign of the estimated factor loadings, higher fitted GFP score is related to more openness, extraversion, agreeableness, conscientiousness, and less neuroticism. As a result, the proposed method reproduces the pattern of the Big Five personality traits given by \cite{funder1991global}. Moreover, the estimated regression coefficients shown
in Figure \ref{regressionCoef} also depict significantly higher GFP score for women as compared to men, and no significant difference between cancer and no cancer groups.

\begin{figure}[tb!]
\begin{minipage}{\dimexpr 0.2\linewidth-3cm\relax}
		\rotatebox{90}{$~$}
		
	\end{minipage}%
	\begin{minipage}{\dimexpr 0.5cm\relax}
		\rotatebox{90}{\scriptsize IFI}
		
	\end{minipage}%
	\begin{minipage}{\dimexpr\linewidth-0.5cm\relax}%
			\begin{minipage}{0.455\linewidth}
				\centering
				\includegraphics[width=\textwidth]{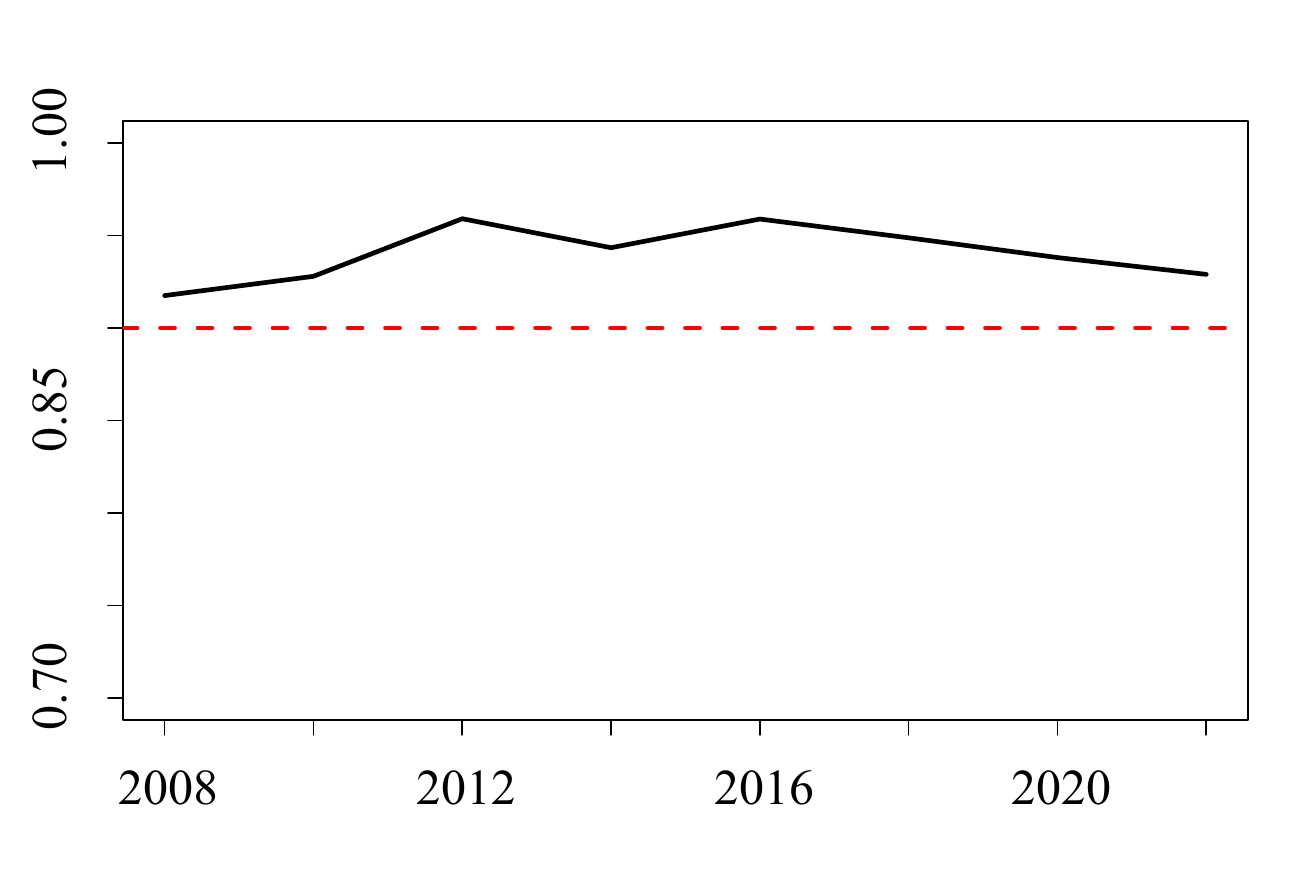}	
			\end{minipage}
   \hspace{-9cm}
			\begin{minipage}{0.63\linewidth}
		\rotatebox{90}{$~$}
		
	\end{minipage}%
			\begin{minipage}{\dimexpr 0.5cm\relax}
		\rotatebox{90}{\scriptsize SRMR}
	\end{minipage}%
			\begin{minipage}{0.45\linewidth}
				\centering
				\includegraphics[width=\textwidth]{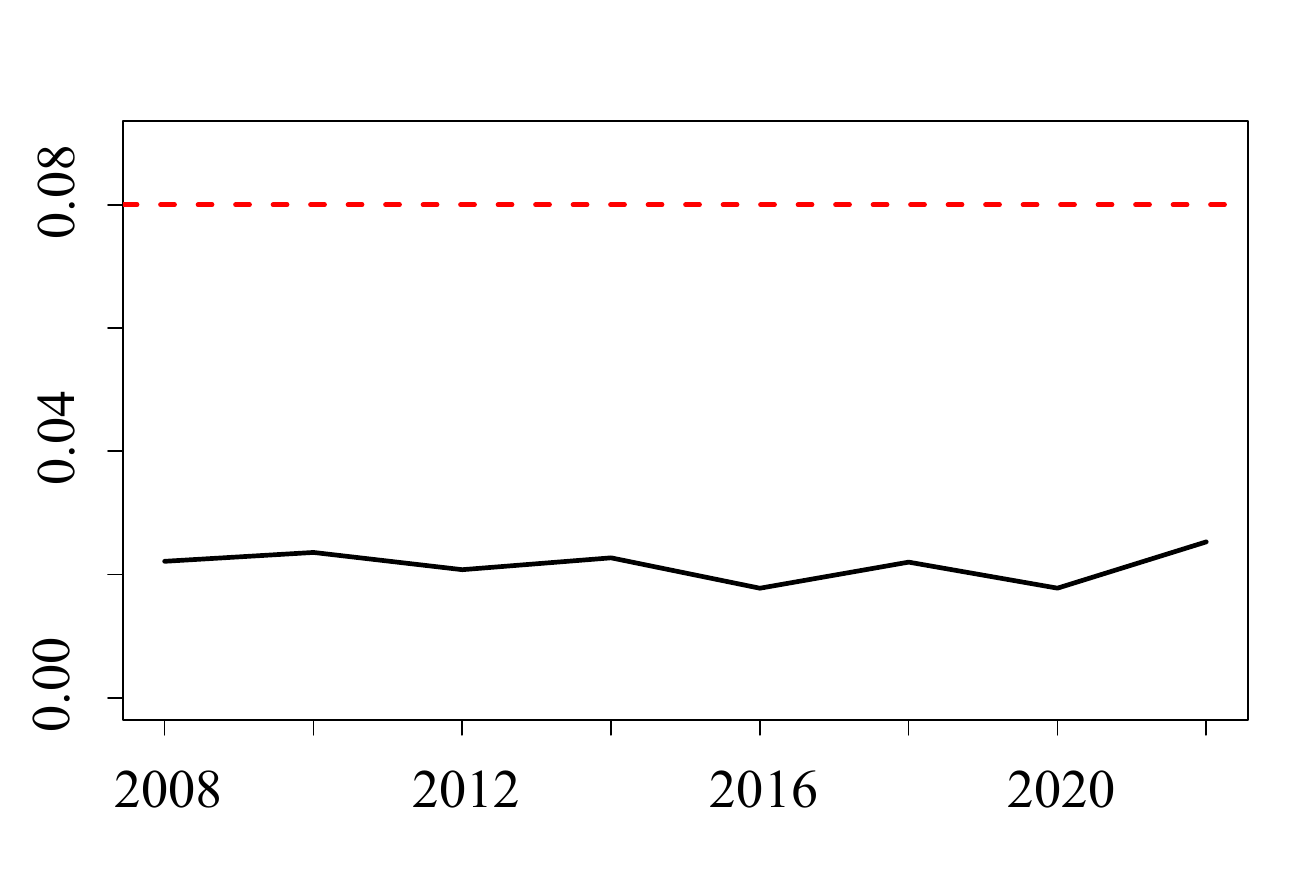} 	
			\end{minipage}
			\begin{minipage}{\dimexpr 0.2\linewidth-1.25cm\relax}
		\rotatebox{90}{$~$}
		
	\end{minipage}%
	\end{minipage}%
	
		\begin{minipage}{1\linewidth}
				\centering
				\hspace{-2em}\footnotesize{Year}
			\end{minipage}
	\caption{Model fit over time. Left panel: IFI function. Right panel: SRMR function.}

	\label{indices}
\end{figure}

To evaluate the FSEM model fit to the data, all the fit indices defined in Section \ref{gofi} were calculated and are presented in Table \ref{inds}. Additionally, the  SRMR and IFI functions for the fit of the model are depicted in Figure \ref{indices}. The values of the indices indicate that the model fit is generally good. Further, according to the SRMR and IFI functions in Figure \ref{indices}, IFI at all time points is higher than 0.9 and SRMR is lower than 0.08, implying the fit is stable and satisfactory over time.

\begin{table}[t]
\centering
	\caption{The values for the goodness of fit indices.}
	\label{inds}
 {\footnotesize
\begin{tabular}{ccccccc}
\vspace{0.05cm}
    $\chi^2/df$ & RMSEA & SRMR  & CFI   & IFI   & GFI   & TLI  \\\hline
 2.549       & 0.094 & 0.022 & 0.922 & 0.940 & 0.956 & 0.893

\end{tabular}
}
\end{table}

\section{Discussion and conclusion}\label{sec:conclusion}
In this paper, we developed a novel more general family of functional SEMs, namely FSEMs, for dealing with observed longitudinal or functional data, and for problems incorporating latent variables that might also possibly be functional. We proposed a maximum likelihood framework that properly handles data with repeated measurements observed at either equally spaced regular or non-equally spaced irregular time points, therefore proposing a framework capable of flexibly handling different missing data structures. We proposed a computational solution based on the EM algorithm to estimate the functional parameters, including the factor loadings, the regression coefficients, the covariance operators of the residual functions, and the unique factors. We provided smooth estimates for the functional parameters by applying a penalized likelihood approach that chooses the smoothing parameters using a cross-validation approach. 

The proposed FSEM model has potentially a very broad impact, as it would be beneficial to researchers from applied fields who need to deal with latent structures in a repeated measurement setting. For example, potential medical applications could include modeling longitudinal trajectories of disease biomarkers as a function of underlying latent disease progression factors. The factor loadings can be chosen as either scalar or functional parameters, depending on the problem at hand: for a measurement model with time-independent loadings one can choose scalar factor loadings, otherwise, to obtain a time-dynamic measurement model, concurrent or historical functional factor loadings can be used.

We examined the performance of the proposed framework thanks to a set of simulation studies considering two different models and various data scenarios. We ran the simulation studies for combinations of different sample sizes, number of time points where the indicator curves are observed, and sampling designs. We considered two sampling designs, one with regular and another with irregular time points, to demonstrate the ability of the proposed framework to perform accurate estimation in the presence of various different data structures. 

The proposed FSEM was also successfully employed to analyze a subset of the data available within the HRS \citep{heeringa1995technical}, to examine the trend over time of the factor loading curves corresponding to the latent variable General Factor of Personality (GFP), and in general to assess whether the results obtained via the FSEM were in accordance to the previous literature on the topic. The relationship between the variable GFP and the covariates `sex' and  `cancer' in the course of eight years from 2008 to 2022 for adults aged 50 and over was also explored thanks to the FSEM fit. The variable GFP was supposed to be identified through its corresponding indicators, including the Big Five personality traits `neuroticism', `extraversion', `openness', `agreeableness', and `conscientiousness'. According to the results derived from fitting the FSEM to the dataset, all the factor loadings of the indicators were positive and constant over time except for the factor loading of `neuroticism', which was negative and changed over time. Relatively constant factor loadings over time support the established idea that the GFP is a stable personal characteristic over time, and also the signs of the estimated factor loadings are in concordance with those reported in the literature. Moreover, there was no significant difference in the GFP score between cancer and no cancer groups, while this variable was higher for women compared to men, both also being quite reasonable and thus expected conclusions. Finally, the model fit was evaluated using goodness of fit indices developed for FSEM, and the values indicated a generally good model fit.

A limitation of the current work concerns the computational scalability to ultra-large datasets of the proposed EM algorithm. Currently, we are limited to handle designs with a few hundreds observations, and in the order of tens factors. However, we are in the process of optimizing the code by using the \texttt{Rcpp} R package \citep{RcppJSS,RcppRpackage}, which will significantly enhance the computational performance. All functions fitting the FSEM, as well as the scripts for reproducing the simulations, are available at \href{https://github.com/asgari-fatemeh/FSEM}{https://github.com/asgari-fatemeh/FSEM}.

We foresee a potential extension for the current work in dealing with the case when the indicators are not fully observed along time, but their discretized versions are observed at each time point, in the same spirit as in \citet{asgari2020latent}. In such a situation, a binary indicator might be simply observed as zero when a certain symptom or condition is not present, and 1 when it is present. A latent functional indicator variable can then be estimated being positive when the corresponding dichotomized indicator is observed to be 1, and negative when it is zero.


Finally, a future research direction could focus on developing a multi-level FSEM for handling clustered or hierarchical data structures. This would extend the current FSEM framework to accommodate smooth functional latent variables and their corresponding smooth functional indicators that vary at different levels of the hierarchy.

\section*{Acknowledgment}
The first author was supported by ERA-NET-NEURON, JTC 2020/Norwegian Research Council (grant number 323047), and the South-Eastern Norway Regional Health Authority (grant number 2021046).

\appendix

\section{Appendix A}\label{ap_a}
\renewcommand{\theequation}{A\arabic{equation}}

 Based on the Karhunen-Lo\`eve expansion  of $\varepsilon_{ij}(t)$, we have 
 \begin{align}\label{ap_a1}
\varepsilon_{ij}=\sum_{r=1}^{J}\sqrt{\nu_{jr}}w_{jr}\phi_{jr}=\boldsymbol{\phi}_{j}^\top\text{diag}\left\{\nu_{j(1)},\ldots, \nu_{j(J)}\right\}^{1/2}\mathbf{w}_{ij},
\end{align}
where the  $J$-dimensional vector $\mathbf{w}_{ij}$ has a multivariate standard Gaussian distribution and $\boldsymbol\phi_{j}$ denotes the vector of eigen-functions for $\varepsilon_{ij}(t)$.
 Suppose the function $y_{ij}(t)$ is evaluated at the time points $t_1,t_2,\ldots,t_M$. From equations (\ref{eq2.3}) and (\ref{ap_a1}), we can write
\begin{align}\label{ap_a2}\nonumber
\begin{bmatrix}
y_{ij}(t_1)  \\
y_{ij}(t_2)  \\
\vdots \\
y_{ij}(t_M) 
\end{bmatrix}=&
\begin{bmatrix}
\beta_{j}(t_1)  \\
\beta_{j}(t_2)  \\
\vdots \\
\beta_{j}(t_M) 
\end{bmatrix}+
\sum_{m=1}^q a'_{jm}
\begin{bmatrix}
f'_{jm}(\eta_{im},t_1)  \\
f'_{jm}(\eta_{im},t_2)  \\
\vdots \\
f'_{jm}(\eta_{im},t_M) 
\end{bmatrix}\\
&+\sum_{m=1}^q a_{jm}
\begin{bmatrix}
f_{jm}(\eta_{im},t_1)  \\
f_{jm}(\eta_{im},t_2)  \\
\vdots \\
f_{jm}(\eta_{im},t_M) 
\end{bmatrix}+
\begin{bmatrix}
\boldsymbol{\phi}_{j}^\top(t_1)  \\
\boldsymbol{\phi}_{j}^\top(t_2)  \\
\vdots \\
\boldsymbol{\phi}_{j}^\top(t_M) 
\end{bmatrix}\mathbf{G}_j^{1/2}\mathbf{w}_{ij}, 
\end{align}
where $\text{diag}\left\{\nu_{j(1)},\ldots, \nu_{j(J)}\right\}$ is denoted by $\mathbf{G}_j$. Using the basis expansions  $y_{ij}(t)=\mathbf{e}^\top(t)\mathbf{y}_{ij}$, $\beta_{ij}(t)=\mathbf{e}^\top(t)\boldsymbol{\beta}_{ij}$, $f_{jm}(\eta_{im},t)=\mathbf{e}^\top(t)\mathbf{f}_{ijm}$, $f'_{jm}(\eta_{im},t)=\mathbf{e}^\top(t)\mathbf{f}'_{ijm}$, and $\boldsymbol{\phi}_j(t)=\boldsymbol{\Phi}_j^\top\mathbf{e}(t)$, where $\left\{ \boldsymbol{\Phi}_{j} \right\}_{rk}= \langle\phi_{j´(k)},e_r\rangle,$ for $r,k=1,\ldots,J$, the equation (\ref{ap_a2}) can be rewritten as
\begin{align}\label{ap_a3}
\mathbf{E}^\top\mathbf{y}_{ij}=\mathbf{E}^\top\boldsymbol{\beta}_{j}+\sum_{m= 1}^{q}\mathbf{f}_{ijm}^{'}a_{jm}^{'}+\sum_{m=1}^q\mathbf{f}_{ijm}a_{jm}+\mathbf{E}^\top\boldsymbol{\Phi}_j\mathbf{G}_j^{1/2}\mathbf{w}_{ij},
\end{align}
where $\mathbf{E}=\left[\mathbf{e}(t_{1}),\mathbf{e}(t_2)\ldots,\mathbf{e}(t_{M})\right]\in \mathbb{R}^{J \times t_M}$ and $\boldsymbol{\Phi}_j\mathbf{G}_j^{1/2}\mathbf{w}_{ij}$ is denoted by $\boldsymbol{\varepsilon}_{ij}$. Finally, by multiplying both sides of the equation (\ref{ap_a3}) by $(\mathbf{E}\mathbf{E}^\top)^{-1}\mathbf{E}$, the model in (\ref{eq4.3}) is derived.

\section{Appendix B}\label{ap_b}
\renewcommand{\theequation}{B\arabic{equation}}

\subsection{Fixed effect}
For the fixed effect, we have $f'_{jm}(\eta_{im},t)=\eta_{im}(t)$ and $f_{jm}(\eta_{im},t)=\lambda_{jm}\eta_{im}(t)$. Based on the basis expansion 
\begin{equation}\label{ap_b1}
    \eta_{im}(t)=\mathbf{e}^\top(t)\boldsymbol{\eta}_{im},
\end{equation}
we can write $\mathbf{f}'_{ijm}=\mathbf{E}^\top\boldsymbol{\eta}_{im}$ and 
$$
\mathbf{f}_{ijm}=\begin{bmatrix}
\boldsymbol{\eta}^\top_{im}\mathbf{e}(t_1)\lambda_{jm}\\
\boldsymbol{\eta}^\top_{im}\mathbf{e}(t_2)\lambda_{jm}\\
\vdots\\
\boldsymbol{\eta}^\top_{im}\mathbf{e}(t_M)\lambda_{jm}
\end{bmatrix}=
(\mathbf{I}_M\otimes\boldsymbol{\eta}_{im})^\top
\begin{bmatrix}
\mathbf{e}(t_1)\\
\mathbf{e}(t_2)\\
\vdots\\
\mathbf{e}(t_M)
\end{bmatrix}\lambda_{jm},
$$
in which $[\mathbf{e}(t_k)]_{1\leq k\leq M}$ is denoted by $\boldsymbol{\omega}$.

\subsection{Concurrent effect}
When the effect is concurrent, we have $f'_{jm}(\eta_{im},t)=\eta_{im}(t)$ and $f_{jm}(\eta_{im},t)=\lambda_{jm}(t)\eta_{im}(t)$. Thus, based on (\ref{ap_b1}) and $\lambda_{jm}(t)=\mathbf{e}^\top(t)\boldsymbol{\lambda}_{jm}$,  we obtain $\mathbf{f}'_{ijm}=\mathbf{E}^\top\boldsymbol{\eta}_{im}$ and 
$$
\mathbf{f}_{ijm}=\begin{bmatrix}
\boldsymbol{\eta}^\top_{im}\mathbf{e}(t_1)\mathbf{e}^\top(t_1)\boldsymbol{\lambda}_{jm}\\
\boldsymbol{\eta}^\top_{im}\mathbf{e}(t_2)\mathbf{e}^\top(t_2)\boldsymbol{\lambda}_{jm}\\
\vdots\\
\boldsymbol{\eta}^\top_{im}\mathbf{e}(t_M)\mathbf{e}^\top(t_M)\boldsymbol{\lambda}_{jm}
\end{bmatrix}=
(\mathbf{I}_M\otimes\boldsymbol{\eta}_{im})^\top
\begin{bmatrix}
\mathbf{e}(t_1)\mathbf{e}^\top(t_1)\\
\mathbf{e}(t_2)\mathbf{e}^\top(t_2)\\
\vdots\\
\mathbf{e}(t_M)\mathbf{e}^\top(t_M)
\end{bmatrix}\boldsymbol{\lambda}_{jm},
$$
where $\big[\mathbf{e}(t_{k})\mathbf{e}^\top(t_{k})\big]_{1\leq k\leq M}$ is denoted by $\boldsymbol{\Omega}_1$.

\subsection{Historical effect}
We know that $f'_{jm}(\eta_{im},t)=\int_{s\leq t}\eta_{im}(s)ds$ and $f_{jm}(\eta_{im},t)=\int_{s\leq t}\lambda_{jm}(s,t)\eta_{im}(s)ds$.  Due to (\ref{ap_b1}), we can write
\begin{equation*}
    \mathbf{f}'_{ijm}=\begin{bmatrix}
\int_{s\leq t_1}\mathbf{e}^\top(s)ds\\
\int_{s\leq t_2}\mathbf{e}^\top(s)ds\\
\vdots\\
\int_{s\leq t_M}\mathbf{e}^\top(s)ds
\end{bmatrix}\boldsymbol{\eta}_{im},
\end{equation*}
where $\left[\int_{s\leq t_{k}}\mathbf{e}^\top(s)ds\right]_{1\leq k\leq M}$ is denoted by $\boldsymbol{\Delta}$. The basis expansion of $\lambda_{jm}(s,t)$ can be written as
\begin{align}\label{ap_b2}\nonumber
\lambda_{jm}(s,t)&=\mathbf{e}^\top(s)[\lambda_{rk}^{(jm)}]_{1\leq r,k\leq J}\mathbf{e}(t)\\
&=(\mathbf{e}^\top(t)\otimes\mathbf{e}^\top(s))\text{vec}([\lambda_{rk}^{(jm)}]_{1\leq r,k\leq J}),   
\end{align}
where $\text{vec}([\lambda_{kh}^{(jm)}]_{1\leq k,h\leq J})$ is denoted by $\boldsymbol{\lambda}_{jm}$. Hence, based on (\ref{ap_b1}) and (\ref{ap_b2}), we have
\begin{align*}
    \mathbf{f}_{ijm}&=\begin{bmatrix}
\boldsymbol{\eta}_{im}^\top\int_{s\leq t_1}\mathbf{e}(s)(\mathbf{e}^\top(t_1)\otimes\mathbf{e}^\top(s))ds\boldsymbol{\lambda}_{jm}\\
\boldsymbol{\eta}_{im}^\top\int_{s\leq t_2}\mathbf{e}(s)(\mathbf{e}^\top(t_2)\otimes\mathbf{e}^\top(s))ds\boldsymbol{\lambda}_{jm}\\
\vdots\\
\boldsymbol{\eta}_{im}^\top\int_{s\leq t_M}\mathbf{e}(s)(\mathbf{e}^\top(t_M)\otimes\mathbf{e}^\top(s))ds\boldsymbol{\lambda}_{jm}
\end{bmatrix}\\
&=(\mathbf{I}_M\otimes\boldsymbol{\eta}_{im})^\top
\begin{bmatrix}
\int_{s\leq t_1}\mathbf{e}(s)(\mathbf{e}^\top(t_1)\otimes\mathbf{e}^\top(s))ds\\
\int_{s\leq t_2}\mathbf{e}(s)(\mathbf{e}^\top(t_2)\otimes\mathbf{e}^\top(s))ds\\
\vdots\\
\int_{s\leq t_M}\mathbf{e}(s)(\mathbf{e}^\top(t_M)\otimes\mathbf{e}^\top(s))ds
\end{bmatrix}\boldsymbol{\lambda}_{jm},
\end{align*}
where $\left[\int_{s\leq t_{k}}\mathbf{e}(s)(\mathbf{e}^\top(t_{k})\otimes\mathbf{e}^\top(s))ds\right]_{1\leq k\leq M}$ is denoted by $\boldsymbol{\Omega}_2$.

\section{Appendix C}\label{ap_c}
\renewcommand{\theequation}{C\arabic{equation}}

\subsection{Smooth effect}
For the smooth effect $s_{ml}^x(x_{il},t)$, let  $\mathbf{g}_l=(g_{l1},g_{l2},\ldots,g_{lJ})$ be a $J$-vector of basis functions of the space $L^2(\tau_{l})$, where $\tau_{l}$ stands for the range of the covariate $x_l$ for all $l=1,2,\ldots,Q$. Then, the basis expansion for $s_{ml}^x(x_{il},t)$ can be written as
\begin{align}\label{ap_c1}\nonumber
s_{ml}^x(x_{il},t)&=\mathbf{g}_l^\top(x_{il})[\gamma^{(l)}_{rk}]_{1\leq r,k\leq J}\mathbf{e}(t)\\
&=(\mathbf{e}^\top(t)\otimes\mathbf{g}^\top_l(x_{il}))\text{vec}\big([\gamma^{(l)}_{rk}]_{1\leq r,k\leq J}\big),
\end{align}
where $\text{vec}\big([\gamma^{(l)}_{rk}]_{1\leq r,k\leq J}\big)$ is denoted by $\boldsymbol{\gamma}_{ml}^x$. Using (\ref{ap_c1}), we can write
\begin{align*}
    \mathbf{s}_{iml}^x=\begin{bmatrix}
      (\mathbf{e}^\top(t_1)\otimes\mathbf{g}^\top_l(x_{il}))\boldsymbol{\gamma}_{ml}^x\\
      (\mathbf{e}^\top(t_2)\otimes\mathbf{g}^\top_l(x_{il}))\boldsymbol{\gamma}_{ml}^x\\
      \vdots\\
      (\mathbf{e}^\top(t_M)\otimes\mathbf{g}^\top_l(x_{il}))\boldsymbol{\gamma}_{ml}^x
    \end{bmatrix}=\Big(\begin{bmatrix}
      \mathbf{e}^\top(t_1)\\
      \mathbf{e}^\top(t_2)\\
      \vdots\\
      \mathbf{e}^\top(t_M)
\end{bmatrix}\otimes\mathbf{g}^\top_l(x_{il})\Big)\boldsymbol{\gamma}_{ml}^x,
\end{align*}
in which $[\mathbf{e}^\top(t_k)]_{1\leq k\leq M}^\top$ is denoted by $\mathbf{E}$.

\subsection{Linear effect}
In case the effect is linear, we know $s_{ml}^x(x_{il},t)=\gamma^x_{ml}(t)x_{il}$. Thus, according to the basis expansion $\gamma^x_{ml}(t)=\mathbf{e}^\top(t)\boldsymbol\gamma^x_{ml}$, we obtain
\begin{align*}
    \mathbf{s}_{iml}^x=\begin{bmatrix}
      \mathbf{e}^\top(t_1)\boldsymbol{\gamma}_{ml}^x x_{il}\\
      \mathbf{e}^\top(t_2)\boldsymbol{\gamma}_{ml}^x x_{il}\\
      \vdots\\
      \mathbf{e}^\top(t_M)\boldsymbol{\gamma}_{ml}^x x_{il}
    \end{bmatrix}=\begin{bmatrix}
      \mathbf{e}^\top(t_1)\\
      \mathbf{e}^\top(t_2)\\
      \vdots\\
      \mathbf{e}^\top(t_M)
\end{bmatrix}\boldsymbol{\gamma}_{ml}^x x_{il}.
\end{align*}

\section{Appendix D}\label{ap_d}
\renewcommand{\theequation}{D\arabic{equation}}

Let us define the  vector of observed variables as 
$$
\begin{bmatrix}
      z_1(t)\\
      z_2(t)\\
      \vdots\\
      z_p(t)\\
      x_{1}(t)\\
      x_2(t)\\
      \vdots\\
      x_Q(t)\\
\end{bmatrix}
$$
According to the models given in (\ref{eq2.3}), (\ref{eq2.3c}), and (\ref{eq3.1}), the elements of the model covariance matrix $\boldsymbol\Sigma_t(\boldsymbol\theta)$ can be derived as
\begin{align}\label{ap_d1}
\text{cov}(z_{j}(t),z_{k}(t))
=\text{cov}(y_j(t),y_k(t))+\text{cov}(\epsilon_{jt},\epsilon_{kt}),\qquad j,k=1,2,\ldots,p,  
\end{align}
\begin{align}\label{ap_d2}
\text{cov}(z_{j}(t),x_{l}(t))
=\text{cov}(y_j(t),x_{l}(t)),\qquad j=1,2,\ldots,p, l=1,2,\ldots,Q,  
\end{align}
and $\text{cov}(x_{r}(t),x_{s}(t))$ for $r,s=1,2,\ldots,Q$.

For the equation (\ref{ap_d1}), we have $\text{cov}(\epsilon_{jt},\epsilon_{kt})=\sigma^2_j$ for $j=k$, otherwise equals zero. Moreover, from (\ref{eq2.3}), we obtain
\begin{align*}
 \text{cov}(y_j(t),y_k(t))=\sum_{m=1}^q\text{cov}\big(f_{jm}(\eta_{m},t),f_{km}(\eta_{m},t)\big)a_{jm}a_{km}+\text{cov}(\varepsilon_{j}(t),\varepsilon_{k}(t)),  
\end{align*}
where $\text{cov}(\varepsilon_{j}(t),\varepsilon_{k}(t))=k_j^\varepsilon(t,t)$ for $j=k$, otherwise is zero. For the concurrent effect, we have
\begin{align*}
 \text{cov}\big(f_{jm}(\eta_{m},t),f_{km}(\eta_{m},t)\big)=\lambda_{jm}(t)\lambda_{km}(t)\text{var}(\eta_{m}(t)),   
\end{align*}
and for the historical effect, we can write
\begin{align*}
 \text{cov}\big(f_{jm}(\eta_{m},t),f_{km}(\eta_{m},t)\big)=\int_{s_1\leq t}\int_{s_2\leq t}\lambda_{jm}(s_1,t)\lambda_{km}(s_2,t)\text{cov}(\eta_{m}(s_1),\eta_{m}(s_2))ds_1ds_2.
\end{align*}
Further, based on equation (\ref{eq3.1}), we can write
\begin{align*}
\text{cov}(\eta_{m}(s_1),\eta_{m}(s_2))=&\sum_{n=1}^q\text{cov}\big(s_{mn}^\eta(\eta_n,s_1),s_{mn}^\eta(\eta_n,s_2)\big)b_{mn}^{\eta^2}\\
&+\sum_{l=1}^Q\text{cov}\big(s_{ml}^x(x_l,s_1),s_{ml}^x(x_l,s_2)\big)b_{ml}^{x^2}+k^\zeta_m(s_1,s_2).    
\end{align*}

For the equation given in (\ref{ap_d2}), we have
\begin{align*}
    \text{cov}(y_j(t),x_l(t))=\sum_{m=1}^q\text{cov}(f_{jm}(\eta_m,t),x_l(t))a_{jm},
\end{align*}
in which
\begin{equation*}
 \text{cov}(f_{jm}(\eta_m,t),x_l(t))=\lambda_{jm}(t)\text{cov}(\eta_m(t),x_l(t)),   
\end{equation*}
in case the effect is concurrent and 
\begin{equation*}
 \text{cov}(f_{jm}(\eta_m,t),x_l(t))=\int_{u\leq t}\lambda_{jm}(u,t)\text{cov}(\eta_m(u),x_l(t))du,   
\end{equation*}
if the effect is historical. Moreover,
\begin{align*}
 \text{cov}(\eta_m(u),x_l(t))=&\sum_{n=1}^q\text{cov}(s^\eta_{mn}(\eta_n,u),x_l(t))b^\eta_{mn}\\
 &+\sum_{r=1}^Q\text{cov}(s^x_{ml}(x_r,u),x_l(t))b^x_{mr}.
\end{align*}

\bibliographystyle{biometrika}
\bibliography{paper-ref}

\end{document}